\documentclass{aanda} 
\usepackage[varg]{txfonts}

\usepackage{graphicx}
\usepackage{amsmath}
\usepackage{amssymb}
\usepackage{natbib}
\usepackage{caption}
\usepackage{tablefootnote}
\usepackage{soul} 
\usepackage[colorlinks = true,
 linkcolor = blue,
 urlcolor = blue,
 citecolor = blue,
 anchorcolor = blue, 
 draft = false]{hyperref}

\usepackage[dvipsnames]{xcolor}
\usepackage{xcolor}

\newcommand{\comment}[1]{}
\defcitealias{Nakoneczny2019}{N19}

\begin{document} 
\title{KiDS-SQuaD II: Machine learning selection of bright extragalactic objects to search for new gravitationally lensed quasars}
 \author{Vladislav Khramtsov$^{1}$, Alexey Sergeyev$^{1,2}$, Chiara Spiniello$^{3,4}$, Crescenzo Tortora$^{5}$, Nicola R. Napolitano$^{3,6}$,\\  Adriano Agnello$^{7}$, Fedor Getman$^{3}$, Jelte T. A. de Jong$^{8}$, Konrad Kuijken$^{9}$, Mario Radovich$^{10}$, \\
 HuanYuan Shan$^{11}$, and Valery Shulga$^{12,2}$}
 \institute{
 $^{1}$ Institute of Astronomy, V. N. Karazin Kharkiv National University, 35 Sumska Str., Kharkiv, Ukraine \\
$^{2}$ Institute of Radio Astronomy of the National Academy of Sciences of Ukraine \\
$^{3}$ INAF - Osservatorio Astronomico di Capodimonte, Salita Moiariello, 16, I-80131 Napoli, Italy \\
$^{4}$ European Southern Observatory, Karl-Schwarschild-Str. 2, 85748 Garching, Germany \\
$^{5}$ INAF -- Osservatorio Astrofisico di Arcetri, Largo Enrico Fermi 5, 50125, Firenze, Italy \\
$^{6}$ School of Physics and Astronomy, Sun Yat-sen University, 2 Daxue Road,  Tangjia, Zhuhai, Guangdong 519082, P.R. China \\
$^{7}$ DARK, Niels Bohr Institute, Copenhagen University,  Lyngbyvej 2, 2100 Copenhagen, Denmark \\
$^{8}$ Kapteyn Astronomical Institute, University of Groningen, PO Box 800, 9700 AV Groningen, the Netherlands \\
$^{9}$ Leiden Observatory, Leiden University, P.O.Box 9513, 2300RA Leiden, The Netherlands \\
$^{10}$ INAF - Osservatorio Astronomico di Padova, via dell'Osservatorio 5, I-35122 Padova, Italy \\
$^{11}$ Shanghai Astronomical Observatory (SHAO), Nandan Road 80, Shanghai 200030, China\\
$^{12}$ College of Physics of Jilin University, Qianjin Street 2699, Changchun, 130012, P.R.China\\
\email{[vld.khramtsov, alexey.v.sergeyev, chiara.spiniello]@gmail.com}
\
 }

\titlerunning{KiDS-SQuaD II}
\authorrunning{Khramtsov, Sergeyev, Spiniello et al.}

 \date{Submitted on June 03, 2019 }
 \abstract
  {The KiDS Strongly lensed QUAsar Detection project (KiDS-SQuaD) aims at finding as many previously undiscovered gravitational lensed quasars as possible in the Kilo Degree Survey. This is the second paper of this series where we present a new, automatic object classification method based on machine learning technique.}
   {The main goal of this paper is to build a catalogue of bright extragalactic objects (galaxies and quasars), from the KiDS Data Release 4, with a minimum stellar contamination, preserving the completeness as much as possible, to then apply morphological methods to select reliable gravitationally lensed quasar candidates.}
   {After testing some of the most used machine learning algorithms, decision trees based classifiers, we decided to use CatBoost, that was specifically trained with the aim of creating a sample of extragalactic sources as clean as possible from stars. We discuss the input data, define the training sample for the classifier, give quantitative estimates of its performances, and finally describe the validation results with \textit{Gaia} DR2, AllWISE, and GAMA catalogues. }
   { We have built and make available to the scientific community the KiDS Bright EXtraGalactic Objects catalogue (KiDS-BEXGO),  specifically created to find gravitational lenses. This is made of $\approx6$ millions of sources classified as quasars ($\approx 200\,000$) and galaxies ($\approx 5.7$M), up to $r<22^m$. From this catalog we selected 'Multiplets': close pairs of quasars or galaxies surrounded by at least one quasar, presenting the 12 most reliable gravitationally lensed quasar candidates, to demonstrate the potential of the catalogue, which will be further explored in a forthcoming paper. We compared our search to the previous one, presented in the first paper from this series, showing that employing a machine learning method decreases the stars-contaminators within the gravitationally lensed candidates.}
   {Our work presents the first comprehensive identification of bright extragalactic objects in KiDS DR4 data, which is for us the first necessary step to find strong gravitational lenses in wide-sky photometric surveys, but has also many other more general astrophysical applications.}
   
\maketitle

\medskip
\section{Introduction}
Strong gravitationally lensed quasars are very rare objects, especially in the case of quadruply lensed \citep{Oguri10}. 
However, it was clear, since the first discovery \citep{Walsh1979}, that these systems are extremely useful tools for observational cosmology, cosmography and extragalactic astrophysics. 

When the light coming from a distant quasar intercepts a massive galaxy, it gets blended and it forms multiple images of the same source, which are often also magnified, becoming brighter. The light-curves of these different images have different paths and thus are offset by a measurable time-delay that depends on the cosmological distances between the observer, the lens and the source, and on the gravitational potential of the lens \citep{Refsdal64}. This time-delays return a one-step measurement of the expansion history of the Universe (primarily $H_0$), and also allow to set constrains on the dark matter halo of the lens galaxy \citep{Suyu14}. 

Moreover, on top of the deflection caused by the lens, the light of the quasar can also be deflected by the gravitational field of other low-mass bodies moving along the line-of-sight ($10^{-6}<m/M_{\odot}<10^{3}$, e.g., single stars, brown dwarfs, planets, globular clusters, etc.). 
This phenomenon, known as microlensing, can be very effective to study the inner structure of the source \citep{Anguita08,Motta12,Sluse11, Guerras2013, Braibant14}, to estimate the masses of these compact bodies \citep{Kochanek04} or to study the stellar content of the lens galaxies \citep{Schechter02, Bate11, Oguri14}. 

Unfortunately, in all these mentioned cases, and especially for cosmography, the biggest limitation lies in the relatively small number of confirmed lenses.  

Thus, taking advantage from the high spatial resolution ($0.2\arcsec$/pixel, \citealt{VST1}) and stringent seeing constraints ($<0.8\arcsec$ in $r$-band) of the Kilo Degree Survey (KiDS, \citealt{kids_main, KiDS3, KiDS4}), we have recently started the KiDS Strongly lensed QUAsar Detection project, KiDS-SQuaD, presented in \citealt{Spiniello18}, hereafter Paper I. 
We are carrying on a systematic census of lensed quasars with the final goal of building a statistically relevant sample of lenses, covering a wide range of parameters (geometrical configurations, deflector masses and morphologies, redshifts and nature of the sources) to study the the dark matter halos of lens galaxies up to ${\rm z}\sim1$ \citep[]{Schechter02, Bate11, Suyu14} as well as the QSO-host galaxy co-evolution up to ${\rm z}\sim2$ \citep[e.g., ][]{Ding2017}, to put constraints on the inner structure of the quasar accretion disk \citep[size and thermal profile; e.g. ][]{Anguita08,Motta12} as well as the broad-line-region geometry \citep[e.g., ][]{Sluse11,Guerras2013,Braibant14} and finally for precise cosmography (e.g \citealt{Suyu17}). 

The first step to find gravitationally lensed quasars is, obviously, to classify objects, selecting quasars and galaxies while minimizing as much as possible the stellar contamination. 

More generally, the identification of extragalactic objects, quasars and galaxies at all redshifts, is a very important task, that can help to answer to a wide range of astrophysical and cosmological questions, such as the relationship between active galactic nuclei (AGN) and host galaxies or the cosmic evolution of Super Massive Black Holes \citep{Kauffmann00,Haehnelt00,Wyithe03, Hopkins06, Shankar09, Shen09} or the formation and evolution of galaxies \citep{Gama1} across cosmic time.

Spectroscopy is without any doubt a powerful way to unambiguously identify and classify extragalactic objects. 
The most comprehensive dataset of spectroscopically confirmed quasars to date is the Sloan Digital Sky Survey \citep{Paris18}, and a few forthcoming spectroscopic surveys will exponentially increase the amount of confirmed quasars, e.g. Dark Energy Spectroscopic Instrument \citep[DESI, ][]{DESI} and 4-meter Multi-Object Spectroscopic Telescope \citep[4MOST, ][]{4most}.
However spectroscopy comes with a price: it is in fact time-expensive or effective only on small patches on the sky. 
Deep wide-field sky photometric surveys, on the other hand, offers nowadays an unprecedented opportunity to carry on this task on a much larger portion of the sky, modulo the development and the use of sophisticated automatic methods (e.g.,  Decision Trees, \citealt{Quinlan86}, Naives Bayes, \citealt{Duda73}, Neural Networks, \citealt{Rumelhart86}, Support Vector Machines, \citealt{vapnik, Cortes95}) to process the very large amount of produced data.  

KiDS, in particular, is the ideal platform to identify and classify objects and, more specifically to search for strong gravitational lenses, because of its excellent (for ground observation standards) seeing quality (mean $r$-band $\approx 0.70$ $FWHM$), deepness ($25^m$ in $r$-band) as well as its wide field of view ($1350$ deg$^2$ have been covered and will be released with DR5). 

The power of KiDS in the objects classification has already been shown by \citealt[hereafter N19,]{Nakoneczny2019} who built and released a catalogue of quasars from KiDS DR3 ($440$ deg$^2$), classified with a random forest supervised machine learning model, trained on Sloan Digital Sky Survey DR14 \citep[SDSS~DR14,][]{SDSS14} spectroscopic data. 
The approach we undertake in this paper is similar to the one presented in \citetalias{Nakoneczny2019}, as we also use KiDS data as input and SDSS as training sample, although we fine-tune and customize our pipelines to be more suitable for the search of lensed quasars. 
Moreover, the biggest difference between these works is that now we have available photometry in nine-bands. 
In fact, the optical data in KIDS are now (starting from DR4) complemented by infrared data from the VISTA Kilo-degree INfrared Galaxy (VIKING) survey, covering the same KiDS area in the $Z,Y,J,H,K_s$ near-infrared bands \citep{VIKING}. 
Thus, the KiDS$\times$VIKING photometric dataset provides a unique deep, wide coverage in nine bands ($u,g,r,i,Z,Y,J,H,K_s$) which has been proved to be extremely effective to separate quasars from stars using photometric characteristics (e.g. \citealt{Carrasco15}).

Indeed, one of the limitations of the first paper of this series was the manual optical colors selection we used to select quasars-like objects. In fact, in this way, the number of final lensed quasar candidates highly depends on the (somehow arbitrary and often calibrated on previous finding) selection criteria. Moreover, generally this number is of the order of $10\div30$ per deg$^{-2}$, making the necessary second step of visual inspection very difficult and long.
Thus, to make our research suitable to deal with the larger amount of data coming from the fourth (and in the future the fifth) KiDS Data Release \citep{KiDS4} and also new deep wide-field surveys, e.g. Euclid \citep{euclid} or LSST \citep{lsst}, here, in this second paper, we developed a method based on machine learning (ML) and on the combination of VIKING and KIDS data that allow us to more efficiently pinpoint high redshift systems while eliminating as much as possible stellar contamination. 

ML methods are, in fact, very effective in identifying quasars (and more generally, extragalactic sources,\citealt{Eyer05,Ball2006,Elting2008, Kim2011,Gieseke2011,ksz2012,Brescia2015,Carrasco15,Peters15,Krakowski2016,Krakowski2018,Viquar2018,Khramtsov2018,Nolte2019,Bai2019}) with respect to any manual colour cut. 
They allow to explore, with a little human intervention and affordable computing time, large datasets, thus selecting candidates with less stringent pre-selection criteria, maximizing the precision (recovery rate) and minimizing the stellar contamination. 

Recently, a class of specific type of classifiers, the ensembles of decision trees, showed their advantage in the identification of extragalactic sources, and in particular quasars \citep{Ball2006,Carrasco15,Hernitschek2016,Schindler2017,Schindler2018,Sergeyev2018,Jin2019}, also, specifically, as already mentioned, within the KiDS collaboration \citetalias{Nakoneczny2019}.
Moreover, ML methods to search for the strong gravitational lenses already exist, although we note that the large majority of them are build to find galaxy-galaxy lenses rather than lensed quasars using a deep learning approach (e.g., \citealt{Cabanac07, Paraficz16, Lanusse18, Metcalf18, Petrillo17, Petrillo19a, Petrillo19b}; but see also \citealt{Agnello15ml, Ostrovski17, KroneMartins2018, Jacobs19} for lensed quasars). 
Moreover, many of these methods are based on the analysis of imaging data directly, rather than on a catalog level like we do in this paper. 

Nevertheless, we decided to build our own classifier in order to be able to fully customize the characteristics and parameters of the algorithm, also given the required completeness and purity we need for the resulting catalogue. It is of crucial importance for us to build a catalogue of extragalactic objects (not only quasars, since in some case the deflector can give a non-negligible contribution to the light of the whole system or the multiple images can be blended together and thus produce in the KiDS catalogue an 'extended' match rather than many 'point-like' ones), that is as clean as possible from stars, and, at the same time, as complete as possible. 
Thus, developing our own tool and releasing the resulting catalogue is the best possible choice. 

As main result of the novel classification pipeline that, specifically developped for our specific task, we present here the catalogue of Bright EXtraGalactic Objects from KiDS DR4 -- KiDS-BEXGO, which we then use to search for gravitationally lensed quasars, using some of the methods and idea already presented in Paper I. 

\medskip
This paper is organized as follows: in Section~\ref{sec:catalog} we give a general overview on the catalogues and data we use. In Section~\ref{sec:classification} we discuss the method to classify objects and isolate extragalactic ones, using optical and infrared deep photometry, and we introduce and describe our own classification pipeline. 
In Section~\ref{sec:results} we present the result of such a pipeline: the KiDS-BEXGO catalogue, and different validation techniques, based on external data, to test the performance of the classifier. Finally, in Section~\ref{sec:lensing} we focus on 'Multiplets': close pairs of quasars, or galaxies surrounded by at least one quasar (within $5''$), which represent the primary input catalogue for our search for strong gravitational lenses. 
We present our conclusions and future perspectives in Section~\ref{sec:conclusions}. In addition, we present in the Appendix a direct comparison of three different machine learning methods, all based on decision trees.

\section{Data overview}
\label{sec:catalog}
\subsection{The input catalogue from KiDS DR4}
The Kilo-Degree Survey \citep[KiDS, ][]{kids_main} is an European Southern Observatory (ESO) public survey, carried on with the VLT Survey Telescope \citep[VST,][]{VST1,VST2}, that covers $1350$ deg$^2$ on sky in four optical broad-band filters, namely $u,g,r,i$. 
Optical data from KiDS are complemented with data from the VISTA Kilo-Degree Infrared Galaxy Survey \citep[VIKING, ][]{VIKING}, that has already completed the observations in five near-infrared bands ($Z,Y,J,H,K_s$) within the same region of the sky. 
The latest KiDS data release (KiDS DR4 \citealt{KiDS4}), encompasses all the survey tiles ($1006$ deg$^2$ in total) already released in the previous KiDS data releases \citep{KiDS3} with additional tiles covering $\approx 550$ new deg$^2$, thus doubling the area coverage of DR3. In addition, infrared photometric data from VIKING is also included in the KiDS DR4 release for the aperture-matched sources \citep{Kids_VIKING}. Typical magnitude limits for each band are 24.2, 25.1, 25.0, 23.6, 22.7, 22.0, 21.8, 21.1, 21.2 (AB magnitudes, $5\sigma$ in $2''$ aperture), with seeing generally below $1.0''$ in $u,g,r,i,Z,Y,J,H,K_s$ bands \citep{Kids_VIKING,KiDS4}.

We started from the KiDS multi-band DR4 catalogue and selected $\approx 45$M sources, that were detected in the $r$-band, which is the one with the best seeing ($0.7''$), and have a match in each of the other eight bands too. 
However, for the implementation of the classification method presented in this paper, we do not use the full catalogue but we limit to 9\,583\,913 
sources with $r<22^m$, covering $\approx$1000 deg$^2$ in all of the 9 filters with small errors on each magnitudes (we remove all the sources with \texttt{MAGERR\_GAAP}$>1^m$ in each of the band).  
In fact, as in \citetalias{Nakoneczny2019}, we also use  
spectroscopically confirmed objects from the Sloan Digital Sky Survey Data Release 14 \citep[SDSS~DR14,][]{SDSS14} as training sample and therefore we limit our inference to bright objects to avoid any extrapolation to unseen regions in the space of features. 

Throughout this paper we always make use of the Gaussian Aperture and PSF \citep[GAaP, ][]{gaap1, gaap2} magnitudes, corrected for extinction.
Finally, as we describe in more details below, we also use the magnitude-dependent parameter \texttt{CLASS\_STAR} for the objects classification. This was already proven to be a very important feature in \citetalias{Nakoneczny2019}. 

\begin{figure}[t]
\center
\includegraphics[scale=0.3,angle=0]{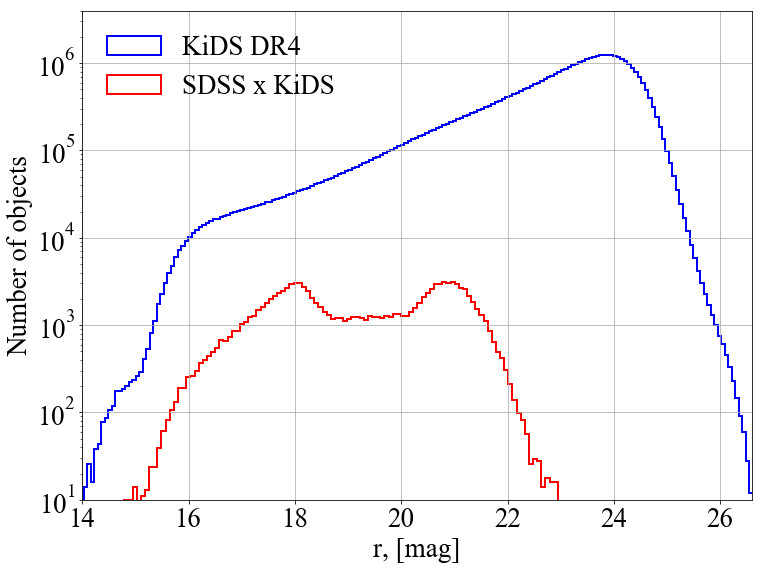}
\caption{Histogram of the $r$ magnitudes for the full KiDS DR4 catalogue (blue) and the training sample from SDSS$\times$KiDS (red).}
\label{fig:rmag}
\end{figure}

The histogram of the $r$-band magnitude distribution for the whole KIDS DR4 and for the spectroscopically confirmed objects that we use as training sample is shown in Figure~\ref{fig:rmag}.
The training sample will be presented in details in the next Section. 

\subsection{The training sample from SDSS DR14}
To provide accurate classification, we need to use a large sample of objects with known true classes. Such data can be obtained from spectroscopic surveys; for our purpose, following the approach of \citetalias{Nakoneczny2019}, we use the SDSS~DR14\footnote{SDSS DR14 is the second release of the Sloan Digital Sky Survey IV phase \citep{sdssiv} and it includes data from all previous SDSS data releases.} catalogue. 
The SDSS DR14 catalogue contains 4\,311\,571 spectroscopically confirmed objects, classified on the basis of their spectra in three main classes: galaxies (2\,546\,963 objects), quasars (824\,548 objects) and stars (940\,060 objects), which we will preserve in our classification setting up a 3-label classification system, as we will describe in details in Section~\ref{sec:classification}. 

We assume that a quasar (hereafter, \texttt{QSO}) is a point-like source\footnote{we note that this assumption has not been made in \citetalias{Nakoneczny2019} that included in their \texttt{QSO} training sample the relatively near ($z<0.2$) AGNs and visible host galaxies.} with \texttt{QSO} class and \texttt{QSO} or \texttt{BROADLINE} subclass; a normal galaxy (hereafter, \texttt{GALAXY}) is an extended source that has a \texttt{GALAXY} class label without \texttt{STARFORMING BROADLINE} and \texttt{STARBURST BROADLINE} subclasses. 
The stars labeling in SDSS does not have subclasses, so simply we assumed that the source is a star (hereafter, \texttt{STAR}) if it has the class \texttt{STAR} from the catalogue.

We cross-match this catalogue with the catalogue of bright sources from KiDS DR4 described above, using a 1.0 arcsec radius, and obtaining, as result, a training sample composed of 183\,048 sources. However some of them have dubious spectroscopic classification.  

A careful cleaning is very important for our scientific purpose, but an automatic masking procedure, eliminating all the dubious cases, that is often applied in classification pipelines to reach the highest possible pureness, is not appropriate here because it might cause the loss of interesting objects with complex morphology and photometric properties, that can be actually good lens candidates\footnote{as a matter of fact, inspecting the misclassified data from SDSS we found three interesting lensed quasar candidates that we selected for spectroscopic follow up. If confirmed, the system will be presented in a forthcoming publication.}. 
We therefore had to pay particular attention to the cleaning procedure which we carried on in a rather manual and interactive way. 
In particular we use this first "unclean" training sample to train the classifiers (which we will describe in the next section). We then visually inspected all the misclassified objects (of the order of few hundreds)\footnote{We use the Navigate SDSS visual tool (\url{http://skyserver.sdss.org/dr14/en/tools/chart/navi.aspx}) to inspect misclassified sources}.  
Interestingly, during the inspection, we discovered, that 
SDSS DR14, indeed, contains few objects with wrong labels, possibly due to a somehow imperfect procedure (among these we found few white dwarfs and few compact galaxies labeled as \texttt{QSO}, blended sources where one of the component is a star, or stars projected into a galaxy) and realized that classifiers trained on such dataset can inherit these mistakes.  
Thus we removed all the the sources, the true class of which did not fit with its imaging and/or spectral properties and we repeat the whole classification pipeline a few times (testing it also with different classifiers, see next section). We note that the total amount of removed sources does not exceed a few percents of the training sample, but that still, the classification results before and after this iterative cleaning procedure are not identical, with the classifier learned with the "clean" training sample producing better results in terms of pureness. 

Finally, another test we did to get a better handle on the importance of our assumptions in building the input catalogue and the best training sample for it, was to change the chosen threshold in the photometric errors of each single band. In particular, we tested three different upper limit for the errors on the magnitudes of the training sample: $1^m$, $0.5^m$ and $0.3^m$. 
As for the cleaning, we trained the classifiers three times with three different training sets made of objects passing these three thresholds and then we compared the performances. We found negligible differences in purity and completeness (at the 0.1\% level) in the classification of the training sample. Then we also compared the predictions for the whole input catalog obtained using the three different training samples, finding, again, no significant differences in the distribution of the sources among the classes. Thus, we decided to use the training sample with the largest number of object and the same error threshold as the input catalogue ($1^m$). 

In conclusion, after removing the sources with 1) bad spectroscopic redshift estimation (for which $\texttt{zWarning}>0$), 2) missing one or more of the 9 optical-infrared magnitudes, 3) high photometric errors ($>1^m$ in each filter) and the 4) accidental duplicates, retrieved after our cross-matching procedure, we ended up with 121\,375 sources, of which 24\,307 sources classified as \texttt{STAR}, 12\,152 sources classified as \texttt{QSO}, and 84\,917 sources  labeled as \texttt{GALAXY}. This catalogue, which we will name for the rest of the paper SDSS$\times$KiDS, will be used in Section~\ref{sec:learning_process} as training sample for the classifiers.  


\section{Classification}
\label{sec:classification}
The main idea behind the classification problem that we have to solve is that it is possible to separate objects into stars, quasars and galaxies on the basics of their photometry, because each family of objects has specific photometric characteristics, which are different from objects belonging to a different family. 
Thus, our first task was to define the feature space where the quasars, galaxies and stars will be located in three different well separated regions. We identify optical-infrared colours as the most suitable features for the objects classification;
since we have 9 magnitudes ($u,g,r,i,Z,Y,J,H,K_s$), there are 36 colours as pairwise differences of various magnitudes. 
We note that this approach is physically-motivated and model-driven. 
\citetalias{Nakoneczny2019} showed that, although magnitudes contribute less to classification than colors and stellarity index, the output based on colors only was different from that using also magnitudes. 
However, we have many more colors at our disposal, thanks to the five additional infrared bands available to us, which make it easier to properly separate stars from quasars and galaxies. Moreover, considering the fact that it is somehow hard to find simple cut in magnitudes that allow to separate different classes of objects, we decided to do not use magnitudes and only consider colors which are more effective in separating the different classes.

Also, we add the \texttt{CLASS\_STAR} flag to the feature set, corresponding to the 'stellarity' of a source and derived from the KiDS $r$-band images, the ones with the best seeing.
This KiDS parameter is a continuous measure of whether the object is extended (\texttt{CLASS\_STAR}=0) or point-like (\texttt{CLASS\_STAR}=1) and has been proved to be a very powerful feature in the classification (e.g. \citetalias{Nakoneczny2019}). 
As showed in \citet{kids_main} (Fig.~8), the \texttt{CLASS\_STAR} parameter depends on the signal-to-noise ratio (SNR) and it is an effective way to separate stars from galaxies only for data with SNR$>50$. Thus, an alternative for selecting the input data to classify, which would probably allow to investigate also fainter magnitudes, might be to put a cut in the SNR rather than on $r$-band magnitude. However, at the present stage, the more severe cut in magnitude is necessary given the limitation of the training sample that we use. 

Colors and stellarity values of the sources correspond to the coordinates in the high-dimensional feature space, in which the classification has been performed. 

As already specified in the introduction, we decided to define a 3-class problem, where the classes correspond to stars, quasars and galaxies. 
In fact, this 3-class labeling  allows us to get the purest identification of quasars, unlike the 2-class (stars and quasars, assuming galaxies as extended sources) or the 4-class (stars, quasars, regular galaxies and galaxies with strong emission lines, e.g. starforming galaxies) schemes.
Also, we stress that a 2-class problem, which only separates stars from extragalactic sources is not enough for our scientific purpose. It is true that to find gravitationally lensed quasars, we need a catalogue that contains both galaxies and quasars, but we need to be able to separate these two classes properly in order to find suitable lens candidates (see Section~\ref{sec:lensing}).

In the following sections we describe the final classification algorithm and calibration strategy that we use to build our catalogue, which was the end product of a large series of tests and experiments we carried out, also using different classification schemes, detailed in the Appendix~\ref{testing_classifiers}. 
In fact, we tested three classifiers based on decision trees (Random Forests and two different Gradient Boosting approaches). We decided, in the end, to use the CatBoost (\citealt{cat1,cat2}), one of the two Gradient Boosting \citep[GB, ][]{gradboost} ensemble algorithms that we tested, because it was the one providing the best performance during the training process, as described in the following Section. 

In general, Gradient Boosting \citep[GB, ][]{gradboost} is an ensemble algorithm that constructs a learner by fitting in an iterative way the gradients of the predictions' residuals of the previously constructed learners, typically decision trees (gradient boosting decision tree, GBDT). 
CatBoost in particular, is a novel, fast, scalable, high performance open-source GBDT library\footnote{https://catboost.ai/}, developed by Yandex researchers and engineers\footnote{https://yandex.com/company/}. 
CatBoost has the great advantage, with respect to other Gradient Boosting algorithms, that it 
uses Ordered Boosting \citep{cat2} to avoid the overfitting problem, as we highlight in the Appendix~\ref{testing_classifiers}.

To our knowledge, this is the first application of the CatBoost algorithm to an astronomical task. 


\subsection{Fine-tuning and learning process}
\label{sec:learning_process}
\begin{figure}[t]
\center
\includegraphics[scale=0.37,angle=0]{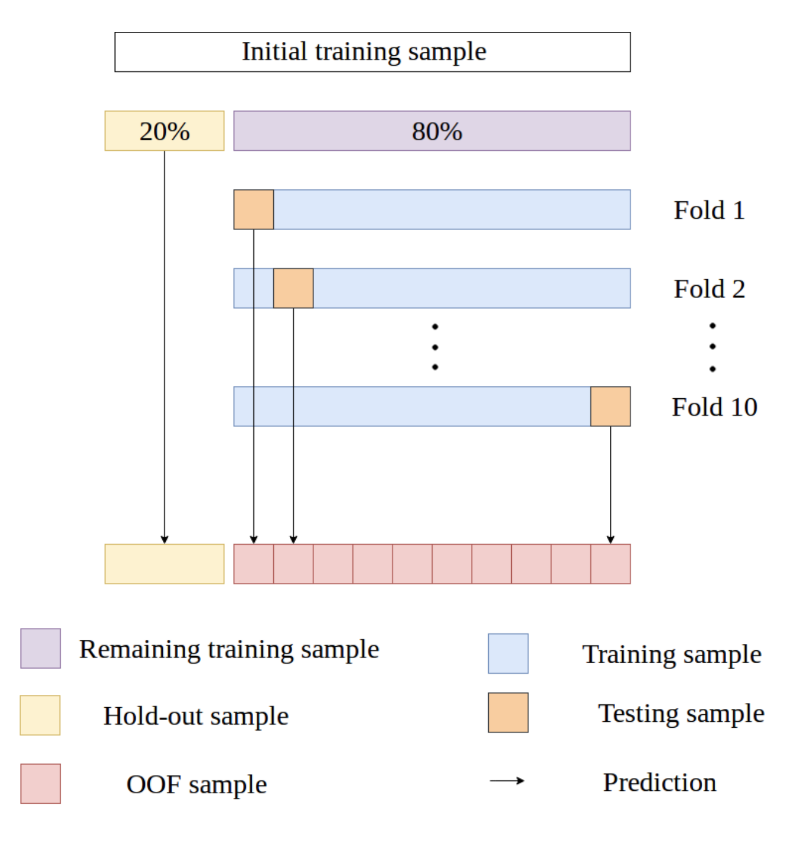}
\caption{A schematic view of the learning via 10-fold cross validation procedure and validation with the OOF and the hold-out samples, drown from the initial training sample}
\label{fig:10cv}
\end{figure}

To be able to analyze the performance of classifier, one need to define a set of validation data and the type of learning with respect to the  training-to-validation division. 
We therefore split the validation into two groups: out-of-fold (OOF) and hold-out. The hold-out sample consists of a random subsample of the initial training data which we keep to access the final performance of the classifiers. 
The remaining part of the initial training sample is used to learn the classifiers with a $k$-fold cross validation procedure. 
This method is one of the most commonly used way to train classifiers and directly compare classification algorithms. Briefly, one divides the training sample into $k$ randomly partitioned disjoint equal parts. 
Then, the classification algorithm trains on $k-1$ parts and the remaining one is used as testing data. This process is repeated $k$ times, each time using one of the $k$ disjoint testing subsamples and obtaining a prediction from it. The combination of these $k$ predictions is the so-called OOF sample. Finally, to obtain the prediction on the new data, one have to make $k$ predictions, from each fold's model, and average them.
A schematic view of the learning process is visualized in Figure~\ref{fig:10cv}. 

Starting from the KiDS$\times$SDSS sample of 121\,376 sources, we randomly selected 20\% of it as hold-out sample and use the remaining 80\% as OOF training sample in the cross-validation process\footnote{The hold-out and OOF samples are kept fixed for all the various algorithms that we tested (see Appendix).} We stress that, among the classifiers that we tested, CatBoost returned the best performance both on the hold-out and OOF samples. 

Before training can take place, the classifier has a list of parameters that have to be tuned to reach the highest possible classification quality. This is true for each of the different classifiers that we tested (see the Appendix for more details on each of them). 
For this purpose, we performed optimal hyperparameter search on 60\% of initial training sample via a 3-fold cross-validation with a a 'BayesSearch' for CatBoost (and XGBoost, while we use a 'GridSearch' method for RF). 

While tuning the wide range of hyperparameters for CatBoost, we noticed that the most influential ones were the \texttt{max\_depth} and the \texttt{early\_stopping} parameters. 
We selected $\texttt{max\_depth}=8$ and $\texttt{early\_stopping}=150$ after BayesSearch, with a maximum number of trees equals $3500$. 

Moreover, 
we applied a weighting criterion to the  loss function for the CatBoost model to further decrease the contamination by stars in the extragalactic objects catalogue (see Appendix~\ref{GBDT} for more details).

After the above described fine-tuning, we finally trained CatBoost with the same training and validation data with a $10$-fold cross-validation (see Fig.~\ref{fig:10cv}). 
The result of the performance for the final CatBoost model (after the fine-tuning) is presented, as confusion matrices, in Figure~\ref{fig:confmatr_end} for the OOF sample (top) and the hold-out sample (bottom). Using the weighting for stars and galaxies, we received a significant improvement in the purity of the quasar sample; in fact, comparing the confusion matrices before weighting loss function (see Appendix) and after that, one can see, that the rate of stars, classified as quasars, decreased from $\approx 0.60\%$ to $\approx 0.30\%$. 
CatBoost lost only $< 1.50\%$ of the quasars, thus only marginally decreasing the resulting completeness of this class. 

\begin{figure}
\includegraphics[scale=0.3,angle=0]{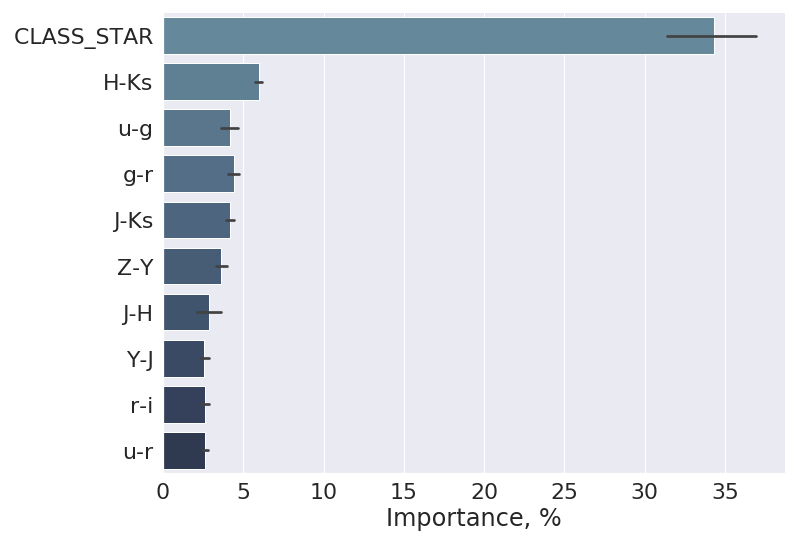}
\caption{Importance of the 10 most significant features, calculated with CatBoost in each of 10 folds. The dispersion of importance for each feature is represented by horizontal ticks at each bar.}
\label{fig:fimportance}
\end{figure}

\begin{figure}
\includegraphics[scale=0.25,angle=0]{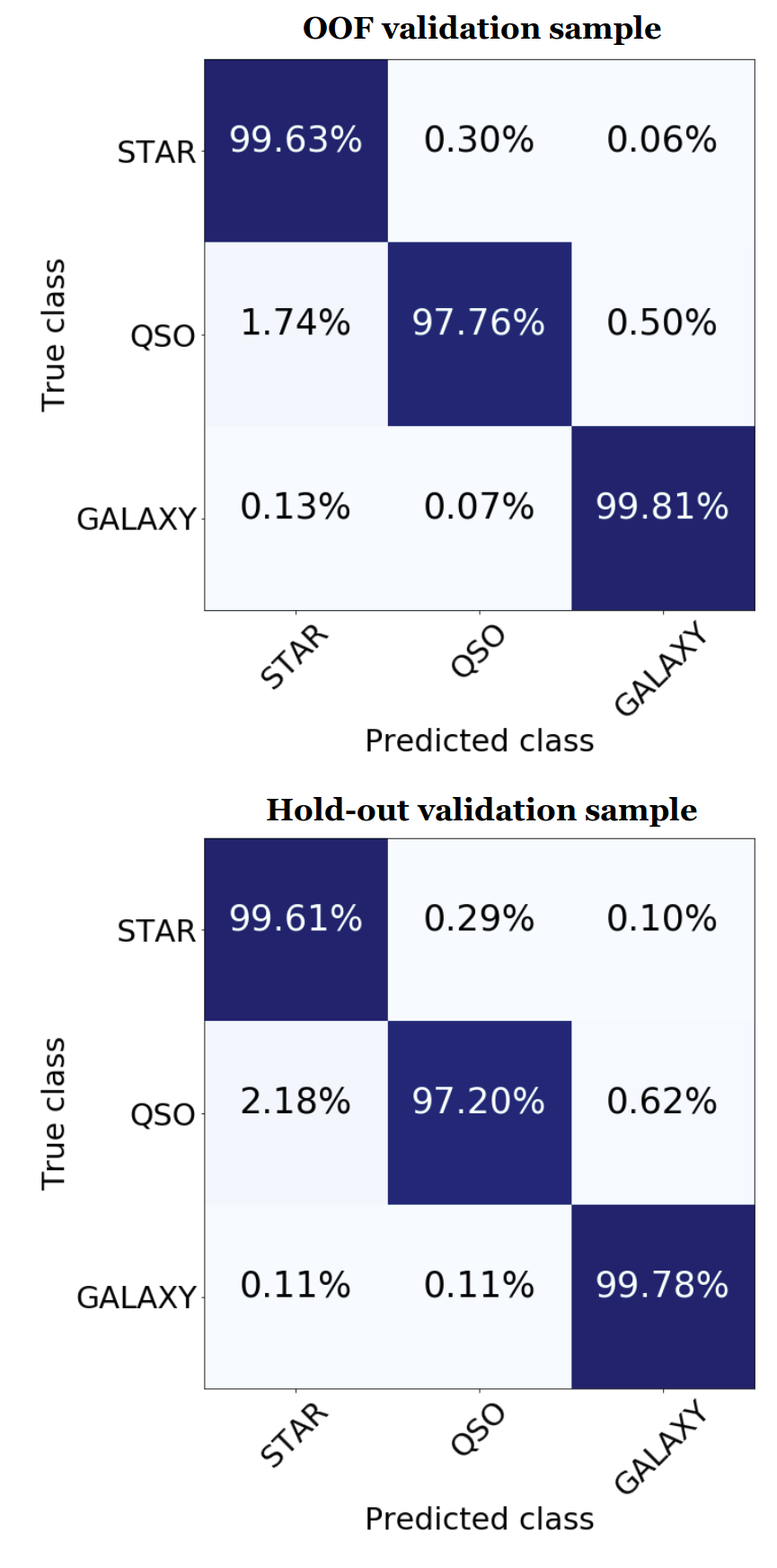}
\caption{Confusion matrices for the final version of CatBoost, after weighting the loss function, performed on the OOF sample (\textit{top} panel) and the hold-out sample (\textit{bottom} panel).}
\label{fig:confmatr_end}
\end{figure}

\begin{figure}[t]
\center
\includegraphics[scale=0.47,angle=0]{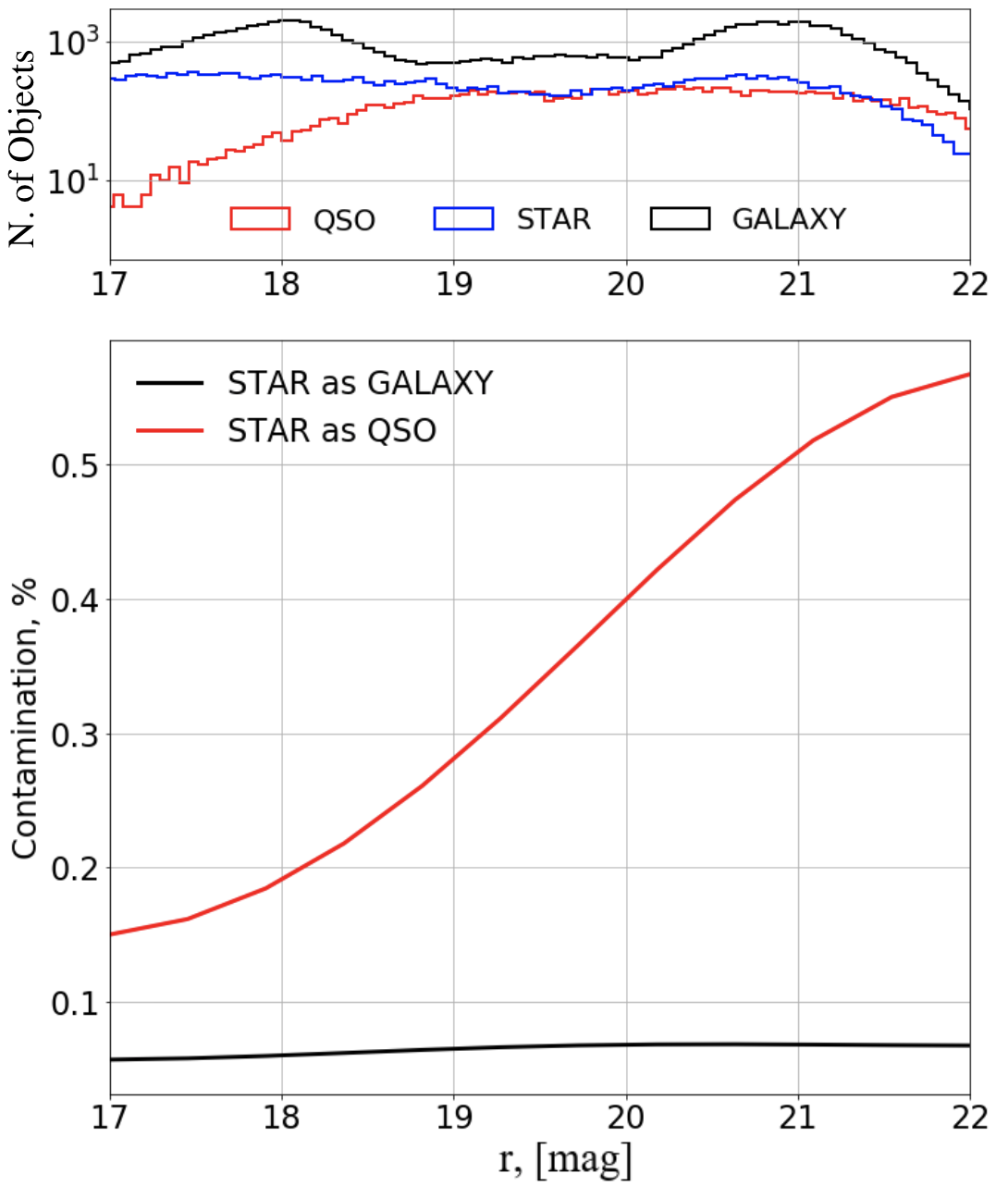}
\caption{Top panel: histogram of the $r$ magnitudes for the three classes of training sources. Bottom panel: the rate of stars, misclassified as quasars (red curve) and galaxies (black curve), as a function of their $r$ magnitude. The plots are produced for the full initial training sample.}
\label{fig:contaminators}
\end{figure}

Another notable result, that we can get with CatBoost, is the relative importance that each feature has for the classification procedure. Feature importance, calculated with decision tree, shows the frequency, with which a certain feature occurs in the tree. In such a way, the higher frequency is directly related to the higher feature contribution to separate the sources, i.e. to the importance of a given feature.  An excellent example of this kind of analysis, together with a full description of the feature importance technique is presented in \citet{DIsanto2018}. 
Figure~\ref{fig:fimportance} shows the 10 most informative features for the all of the CatBoost models, trained one by one via 10-fold cross-validation. Among these, the \texttt{CLASS\_STAR} is certainly the most important one, followed by $H-K_s$, $u-g$, $g-r$, and $J-K_s$. 
This is in perfect agreement with a number of results in the literature, e.g., using $u-g$ and $g-r$ colour diagram it is possible to separate low-redshift quasars from stars \citep{quas_colors1, Carrasco15}; these features, together with the stellarity, are in fact also the most important ones in other ML based classificators (e.g. \citetalias{Nakoneczny2019}). 
Also, quasars at $z\approx2.5$ and $z\approx5.6$ may be recovered by employing $K$-band information in the colour space \citep{quas_colors2}. And finally, it is well known, and also intuitively easy to understand, that morphological information (described here by the \texttt{CLASS\_STAR} feature) allows us to clearly select galaxies, dividing them from stars and quasars in the relatively bright magnitude range that we consider ($r<22^m$).


The maximum rate of stars-contaminators per bin of $r$-band magnitude 
in the quasars catalogue equals $\le0.6\%$ and is expected at the faint end of the sample ($r\approx22$ mag). 
Instead, the stars misclassified as galaxies span over the full optical $r$ magnitude range and does not exceed the $0.1\%$. This is clearly shown in the bottom panel of Figure~\ref{fig:contaminators}). 

Finally, we checked the contamination rate against the signal-to-noise ratio (SNR) in $u$-band, which is the noisiest one, finding that the relative contamination of stars decreases at each magnitude bin by $\sim2\%$ when we only consider objects with SNR $>100$. 

For the input data, whose distribution in the feature space should be similar to that of the training set, we expect a contamination of $0.3\%$ from stars and $0.1\%$ from galaxies in the sample of quasars, and $0.1\%$ from stars and $0.6\%$ from quasars in the sample of galaxies. Thus, we conclude that the algorithm is able to correctly classify up to $97.5\%$ of all the bright quasars from the KiDS DR4 data, and up to $99.8\%$ of the galaxies. 

Surely, these estimations are ideal and do not really reflect the real situation, because they are only based on the training sample, that is a much smaller and simpler case than the full KiDS DR4 catalogue. 
A more realistic estimate of the quality of our extragalactic catalogue, in terms of purity and completeness, can be obtained using the external data to validate the resulting sample of classified sources, as we will do in Section~\ref{sec:catalog}). 

We stress that for our final scientific purpose of finding gravitational lensed quasars, the most crucial point is to be able to get rid of the stellar contaminants. It is, in fact, of fundamental importance to separate as well as possible stars from quasars, being both point-like sources. 

Since the KiDS DR4 input catalogue consists of galaxies mostly, there will be a non negligible number of galaxies contaminating the quasars sample. 
However, as we will explain in more details in Section~\ref{sec:lensing}, strong lenses can be classified as \texttt{GALAXY}, if the deflector gives a non-negligible contribution to the light and the multiple images of quasar are not deblended (thus, the whole system will result in the one extended object with colors which are a mix of galaxies and quasars typical colors), or they can be identified as multiple quasars.
This is the main reason why we build and inspect a catalogue containing all the extragalactic sources (i.e. \texttt{QSO}+\texttt{GALAXY}), looking for 'multiplets'(i.e. sources classified as \texttt{QSO} and with at least one near-by \texttt{QSO} companion) to find lenses belonging to the latter group and by looking for galaxies with at least two quasars near-by (within $5\arcsec$) to find lenses belonging to the former group.

\begin{table}[b]
\caption{Number of resulting objects in the classified KiDS DR4 catalogue, for each class with different probability threshold to define class belonging.}
\centering
\label{p_cut}
\begin{tabular}{ c c c c }
\hline
\hline
 cut-off 	&	 $p_{\texttt{QSO}}$ 	&	 $p_{\texttt{GALAXY}}$	&	 $p_{\texttt{STAR}}$\\
\hline						
$>0.99$ 	&	 62\,425 	&	 5\,538\,193	&	 3\,001\,287\\ 
\hline						
$>0.95$ 	&	 112\,222 	&	 5\,605\,735	&	 3\,533\,787\\
\hline						
$>0.90$ 	&	 128\,393 	&	 5\,629\,623	&	  3\,611\,762\\ 
\hline						
$>0.80$ 	&	 145\,653 	&	 5\,655\,586	&	  3\,660\,368\\ 
\hline						
$>0.67$ 	&	 161\,818 	&	 5\,673\,902	&	  3\,688\,514\\ 
\hline						
$>0.50$ 	&	 181\,336 	&	 5\,690\,885	&	  3\,711\,692\\ 
\hline						

\end{tabular}
\end{table}

\section{The Bright EXtraGalactic Objects Catalogue in KiDS DR4 (KiDS-BEXGO)}
\label{sec:results}
The outputs of the CatBoost classifier for each object are three numbers which represent the probability of belonging to the different classes of objects: $p_{\texttt{STAR}}$, $p_{\texttt{GALAXY}}$, $p_{\texttt{QSO}}$. 

In general, we assume that a source belongs to a given class when the probability of being in that class is the highest. 
With this simple assumption, starting from the input 9.5 million sources in the KiDS DR4 catalogue, we retrieved: 181\,336 quasars, 3\,711\,692 stars and 5\,690\,885 galaxies. 
Using instead a more severe threshold, i.e., considering that an object belongs to a class when the corresponding probability is $>0.8$, we obtain: 5\,665\,586 (59\%) "sure" galaxies, 3\,660\,368 (38\%) "sure" stars and 145\,653 (1.5\%) "sure" quasars, plus 122\,306 objects (1.3\%) with "unsecure" classification. 

We note that for the classification of objects in the final catalog we stick to the original assumption that a source belongs to the class with the largest probability, without applying any further threshold, since "unsecure" extragalctic sources (with $p_{\texttt{GALAXY}}\sim p_{\texttt{QSO}}$) could very well be good lens candidates where the deflector and the quasar images are blended and all give a contribution to the light of the system.

\begin{figure}[t]
\center
\includegraphics[scale=0.38,angle=0]{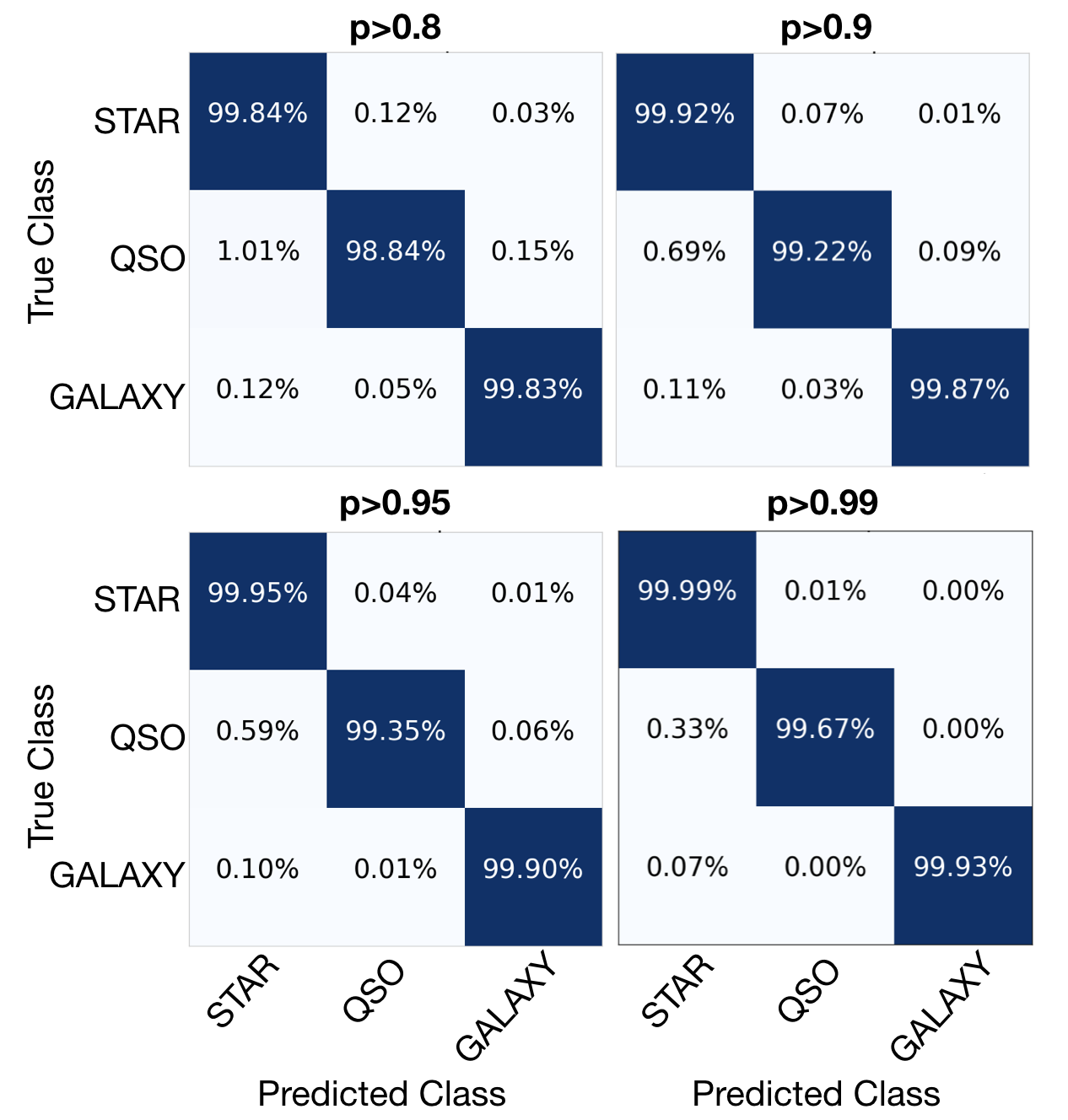}\\
\vspace{0.7cm}
\includegraphics[scale=0.44,angle=0]{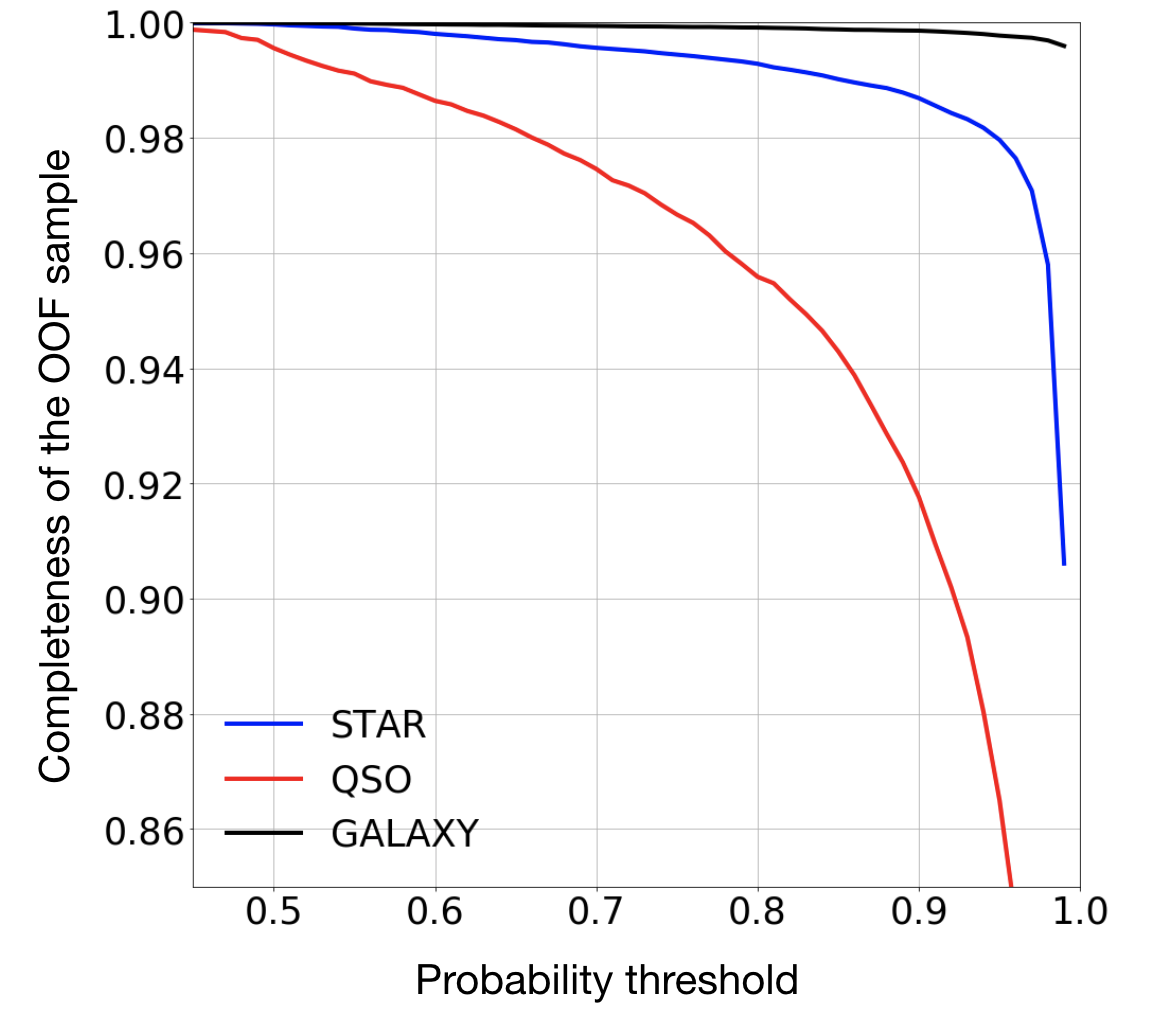}
\caption{The upper panel shows the confusion matrices for the final version of CatBoost performed on the OOF sample using different threshold of probability in the objects classification (see text for more details). 
The lower panel shows instead the completeness rate of the OOF sample as a function of the adopted probability threshold for each class.}
\label{fig:confmatr-pro}
\end{figure}

However, since the levels of completeness and purity depend on the chosen probability, and different scientific cases might require different levels, we provide the number of objects classified in each subsample for 5 different thresholds in Table~\ref{p_cut}.

We also show the confusion matrices obtained for the OOF training sample for the four highest probability levels (0.8, 0.9, 0.95, 0.99) and the completeness rate as a function of the probability threshold for the three classes, in Fig.~\ref{fig:confmatr-pro}, where the thresholds were applied to each of the classes. 

\begin{figure}[t]
\includegraphics[scale=0.28,angle=0]{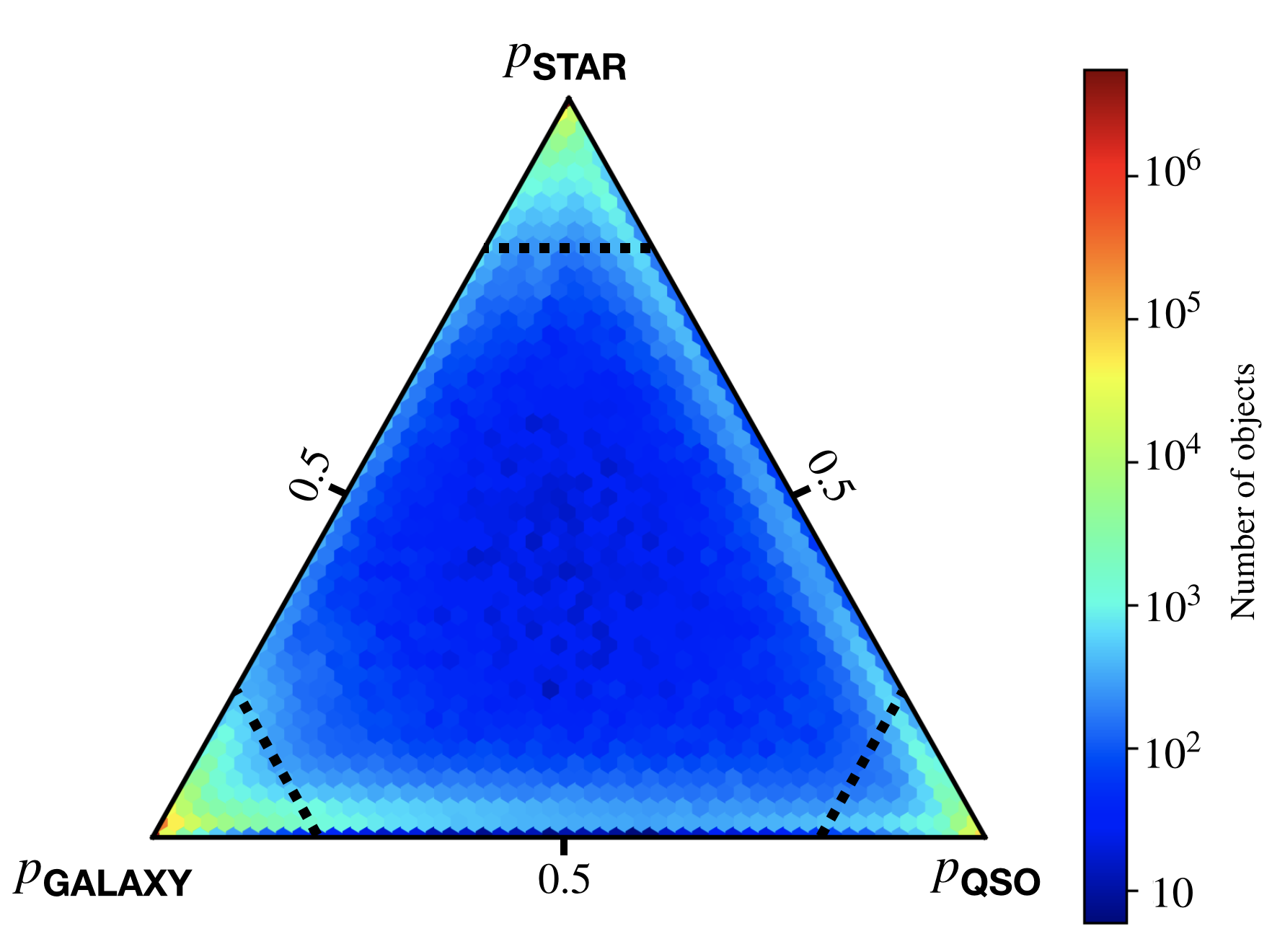}
\caption{Density plot of the final distribution of sources among the classes in the KiDS DR4 catalogue. The triangle corners show the maximum probability to belong to a given family (left \texttt{QSO}, right \texttt{GALAXY}, up \texttt{STAR}), and colors indicate number of objects. Dashed lines correspond to the $p=0.8$ threshold.}
\label{fig:triangle}
\end{figure}

Finally, Fig.~\ref{fig:triangle} provides a visualization of the class distribution of the classified KiDS DR4 catalogue with a density plot, where each corner of the triangle represents the maximum probability to belong to a given class. Objects within the region delimited by dotted lines are "sure", according to the threshold given above ($p>0.8$). 


In the next section, we will only focus on all the objects with $p_{\texttt{QSO}}>p_{\texttt{STAR}}$ or $p_{\texttt{GALAXY}}>p_{\texttt{STAR}}$, that form the Bright EXtraGalactic Objects Catalogue in KiDS DR4 (KiDS-BEXGO) that we will then use in Section~\ref{sec:lensing} for the gravitational lens search. 
We discuss here instead three of the many possible validation procedures, for one or more classes of objects, performed using external data (from the \textit{Gaia} astrometric survey, from the AllWISE infrared catalogue, from the GAMA survey).   
Using external dataset to validate catalogues obtained with ML techniques is a rather standard procedure, as e.g. already shown in \citetalias{Nakoneczny2019} and \citet{Khramtsov18}, although in latter case the PMA \citep{pma} catalogue of proper motions was used to validate purity of galaxies.

Given the results presented in the tests below, together with predictions on the hold-out sample, we are very confident that our ML classifier is able to minimize the stellar contamination in the BEXGO catalogue, which is the first, most crucial step if aiming at digging for gravitationally lensed quasars within very large catalogues.

\begin{figure}[b]
\center
\includegraphics[scale=0.35,angle=0]{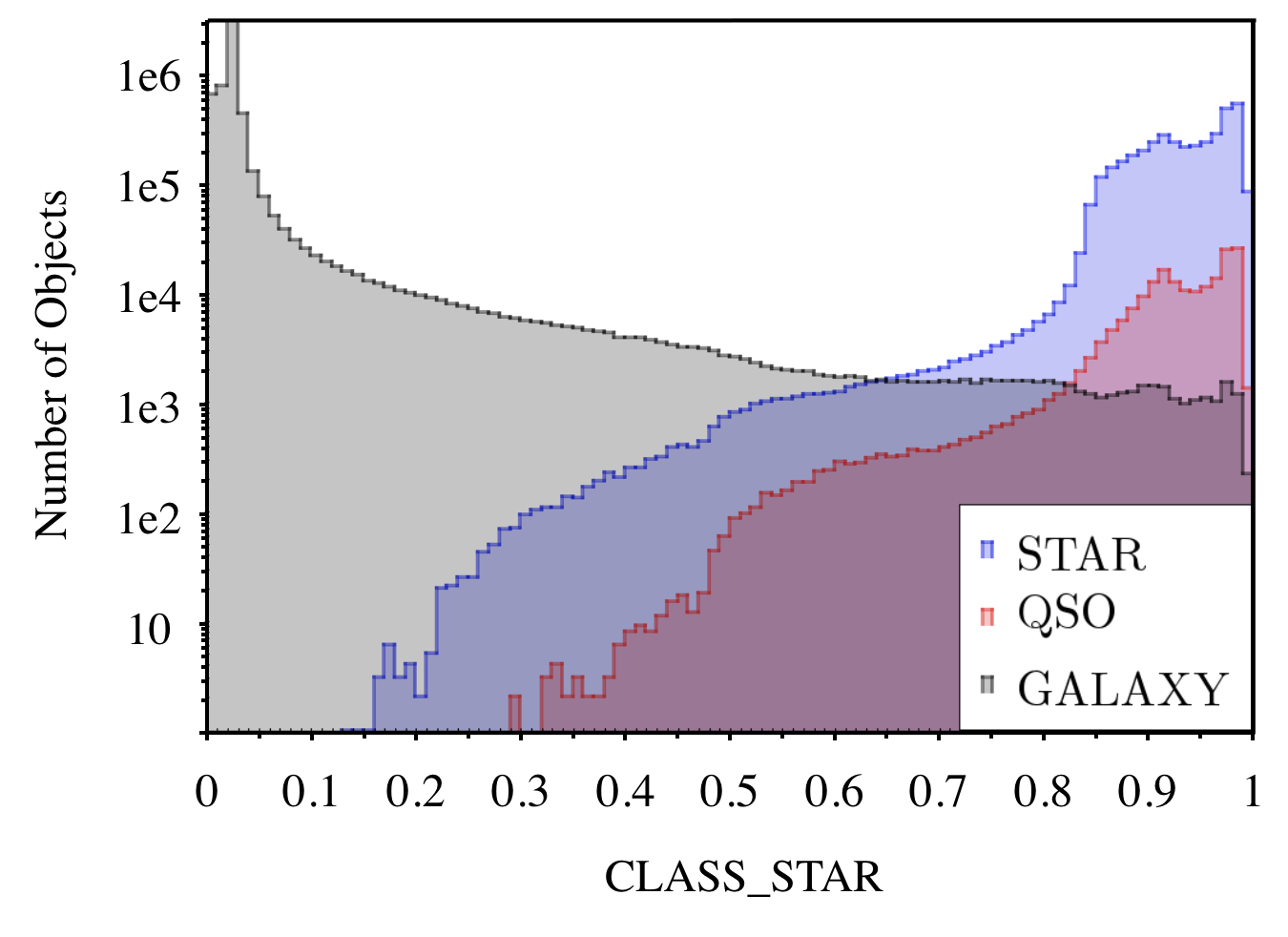}
\caption{Distribution of \texttt{CLASS\_STAR} parameter for the sources from KiDS DR4, classified as galaxies (black), quasars (red) and stars (blue). }
\label{fig:classstar_dr4}
\end{figure}

\begin{figure*}[t]
\center
\includegraphics[scale=0.52,angle=0]{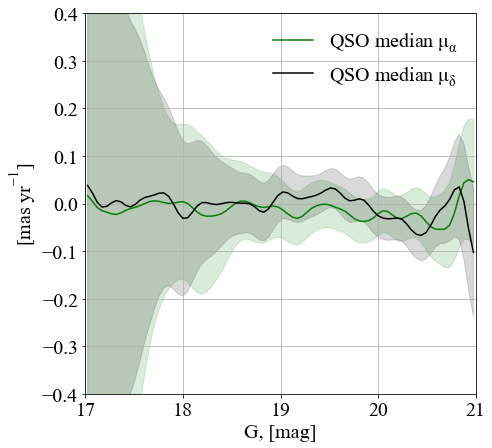}
\includegraphics[scale=0.52,angle=0]{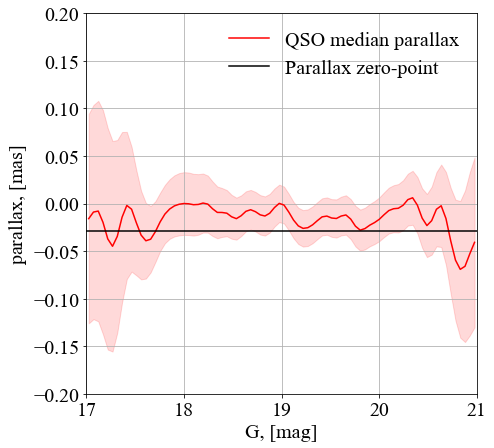}
\caption{Median right ascension (green curve, \textit{left}) and declination (black curve, \textit{left}) proper motion components, and parallaxes (\textit{right}) for the KiDS DR4 \texttt{QSO} sample as a functions of $G$ magnitude. The colored areas represent the standard deviation of the mean $\sigma / N$, where $\sigma$ is the standard deviation and $N$ is the number of quasars in each bin. The black line in the \textit{right} plot represents the parallax zero-point (equals to $-0.029$ mas,  \citealt[][]{gaia_lindegren}).}
\label{fig:gaia}
\end{figure*}

\subsection{Astrometric validation}
\label{sec:validation}
Recently, the \textit{Gaia} \citep[][]{gc1} astrometric survey has provided optical realization of the International Celestial Reference System, materialized with $\approx 500\,000$ quasars, and named \textit{Gaia} DR2 Celestial Reference Frame \citep[\textit{Gaia}-CRF2, ][]{gc2b,gaia_lindegren}. 

The latest data release, \textit{Gaia} DR2 \citep[][]{gc2a}, introduced 5 astrometric parameters (positions $\alpha, \delta$, proper motions $\mu_{\alpha}, \mu_{\delta}$, and parallaxes $\Bar{\omega}$) for $1.3$ billion sources, covering the whole celestial sphere up to $G<21^m$\footnote{This limit corresponds to $r\approx 21^m$ for quasars at $z\leq3$, \citep{proft2015}}. 
The systematical errors in \textit{Gaia} DR2, estimated with a large sample of quasars, do not exceed $0.03$ mas. 
Thus, the \textit{Gaia} DR2 provides an excellent mean for testing the purity of our catalogue, especially for quasars.  
In fact, one of the main observational properties of quasars, that can be used to validate the sample of candidates classified as such, is positional stationary in the optical wavelength range. 
Being quasars very distant sources, they have proper motions of only few microarcseconds, due to different cosmological effects \citep[][]{quasar_pm}. 

We cross-matched the KiDS DR4 sample of 9.5 million classified sources with the \textit{Gaia} DR2 catalogue using a $0''.5$ radius, and retrieved a sample of sources with defined astrometric parameters, of which 52\,636 were classified as \texttt{QSO}, 2\,369\,414 were classified as \texttt{STAR} and 25\,346 -- as \texttt{GALAXY}. 

We checked the proper motions and parallaxes of all the objects classified as quasars and with a match in \textit{Gaia}, to test this assumption that quasars are indeed zero-proper motion and zero-parallax sources within the systematic errors. The results of this test are shown in Figure~\ref{fig:gaia}.

The behaviour of the proper motion components is consistent with the estimated contamination of stars within the quasar subsample of the KiDS DR4 catalogue (Fig.~\ref{fig:contaminators}). 
In fact, at the faintest magnitudes ($20^m.5 < G < 21^m.0$), the proper motion components deviates strongly, due to a larger contamination from stars. At the bright end of magnitudes, the standard deviation of the mean of proper motions and parallaxes is also large, but this is rather due to relatively small amount of sources and, possibly, to contamination from stars. Also, it is important to note, that the parallax (right plot) is biased for the sample of extragalactic sources towards the value of $-0.029$ mas \citep{gaia_lindegren}. 

According to the statistical measures of astrometric parameters, that is reported in Table~\ref{gaia_stats}, we thus can conclude, that the sample of KiDS DR4 quasars mainly consists of motionless sources. There is some disagreement between median and mean values of the parameters, that can be explained by the existence of stars within the sample of quasars with high proper motions (up to $\pm$ 40 mas yr$^{-1}$ at least in one of the components) and parallaxes (up to 35 mas).
A more detailed astrometric analysis providing a more quantitative estimation of the rate of contaminating stars, cannot be produced without accurate modeling and involving another external datasets, which goes beyond the purposes of this paper.

\begin{table}[t]
\caption{Basic statistics of astrometric parameters for KiDS DR4 \texttt{QSO}, cross-matched with \textit{Gaia} DR2}
\centering
\label{gaia_stats}
\begin{tabular}{ c c c c }
\hline
\hline
 Parameter & Mean & Median & Standard deviation\\
\hline
$\Bar{\omega}$, [mas] & -0.010 & -0.014 & 1.125\\
\hline
$\mu_{\alpha}$, [mas yr$^{-1}$] & -0.028 & -0.018 & 2.104\\
\hline
$\mu_{\delta}$, [mas yr$^{-1}$] &  -0.104 & -0.005 & 2.005\\
\hline
\end{tabular}
\end{table}

In order to add something for the galaxies, we use the very simple argument that, by construction, \textit{Gaia} should contains no galaxies at all \citep{Robin2012}. Thus all of the objects with high $p_{\texttt{GALAXY}}$ should not have a match in \textit{Gaia} DR2. This is, of course, only a rough approximation since there might be a number of galaxies that \textit{Gaia} still measures, as for instance, objects with bright cores.

Among the $\approx$25\,000 \texttt{GALAXY} with a match in \textit{Gaia}, we note that only 1\,784 have \texttt{CLASS\_STAR}$>0.5$, thus they can be point-like sources in KiDS, that our algorithm misclassified, or very compact galaxies below the KiDS resolution.

In Figure~\ref{fig:classstar_dr4} we show the distribution of the \texttt{CLASS\_STAR} parameter for each class of full KiDS DR4 catalogue. 
Assuming that galaxies are all extended objects, we would expect to find in KiDS no objects classified as \texttt{GALAXY} with \texttt{CLASS\_STAR}$>0.5$. 

However, there are  objects that are point-like, according to their \texttt{CLASS\_STAR} value, but have been classified as \texttt{GALAXY} by our algorithm. The number of point-like galaxies from Figure~\ref{fig:classstar_dr4} is larger than a couple of thousands, as predicted by the cross-match with \textit{Gaia}. This slight disagreement might be explained by the better resolution of \textit{Gaia} \citep{KroneMartins2018}: these sources might be seen as point-like in KiDS, but are extended and thus not identified in \textit{Gaia}. 

Despite this, the majority of \texttt{GALAXY} sources with a \textit{Gaia} match are indeed extended objects in KiDS, or sources near by a bright star, as we directly verified on a random sample of $\approx$5000 objects, via the SDSS DR14 Navigate Tool\footnote{\url{http://skyserver.sdss.org/dr14/en/tools/chart/navi.aspx}}, and then also checking the KiDS $r$-band images, finding, for most of the sources, bright features (e.g., cores, regions in arms, etc.), that could be resolved only for galaxies with significant angular size.

\subsection{Validation of quasars with mid-infrared data from WISE}
Using mid-infrared (MIR) colours is a very effective way to separate quasars from stars and passive galaxies. In fact, unlike stars and inactive galaxies, that show approximately zero MIR colors, the emission of AGNs conforms to the power-law emission in MIR wavelength range, that causes higher red MIR colours \citep[][]{Elvis1994,stern2005,Assef2013}.

As largely demonstrated by a number of published works, including Paper I, by using a combination of infrared color and magnitudes cuts, it is possible to separate quasars from stars and galaxies (e.g. the two-color criteria of \cite{Lacy2004}, \cite{stern2005}, and \cite{Donley2012} with \textit{Spitzer} \citep{Werner2004} data; or the two-color criteria in \cite{Jarrett2011} and \cite{Mateos2012} or the one-color criteria of \cite{stern2012} and \cite{Assef2013} using data from the Wide-field Infrared Survey Explorer (WISE, \citealt{Wright2010}).

Here, we decided to use the single infrared one-colour cut: [3.6]$\mu$m-[4.5]$\mu$m$ > 0.8$ proposed by \citet{stern2012}, using data from WISE, the NASA space mission, aimed to map all sky in 4 MIR bands: $W1,W2,W3,W4$ (3.6,4.5,12 and 22 $\mu$m respectively). 
This criterion can separate quasars with resulting purity of $\approx 95\%$, but allows one to select quasars only up to $z\approx3.5$ \citep{Guo2018}. We caution the reader that a given sample selected with this criterion can be contaminated by brown dwarfs, that have similar colours. The more elaborated two-colour criterion of \cite{Mateos2012} allows to reduce this contamination, but it requires reliable measurements in the [12]$\mu$m band, which would significantly decreases the total number of matched sources in our case.

We note that, in general, it is harder to validate the purity of galaxies in the same way since stars overlap with (non-active) galaxies in this dimension (see, e.g. Fig.~12 in \citealt{Wright2010}).  

Finally, we clarify to the reader that in this paper, the WISE data is only used as validation for the quasars catalogue but not for the lenses search. In fact, in Paper I, we highlighted that the bottle-neck of our search was indeed the too severe colour WISE pre-selection. 
This could be caused by the fact that, in case the lens and the source are blended in WISE and the deflector gives a large contribution to the light, the colours of this {\sl effective} source may be not quasar-like anymore and move indeed toward lower $W1-W2$ values. Here we rely on a much solid and trustable way to classify objects, our ML based classifier, and thus we do not need to apply any cut nor we need to require a match with WISE to build our candidate list.

We cross-matched the SDSS training sample as well as the catalogue of all the classified objects (Section~\ref{sec:catalog}) with the AllWISE \citep{Cutri2013} data release using a $2''.0$ radius. 
The resulting sample consists of 114\,773 quasars, 3\,289\,858 galaxies, and 2\,020\,768 stars for the classified objects and of 8\,879 quasars, 78\,816 galaxies, and 13\,249 stars for the training data. 

\begin{figure*}
\center
\includegraphics[scale=0.3,angle=0]{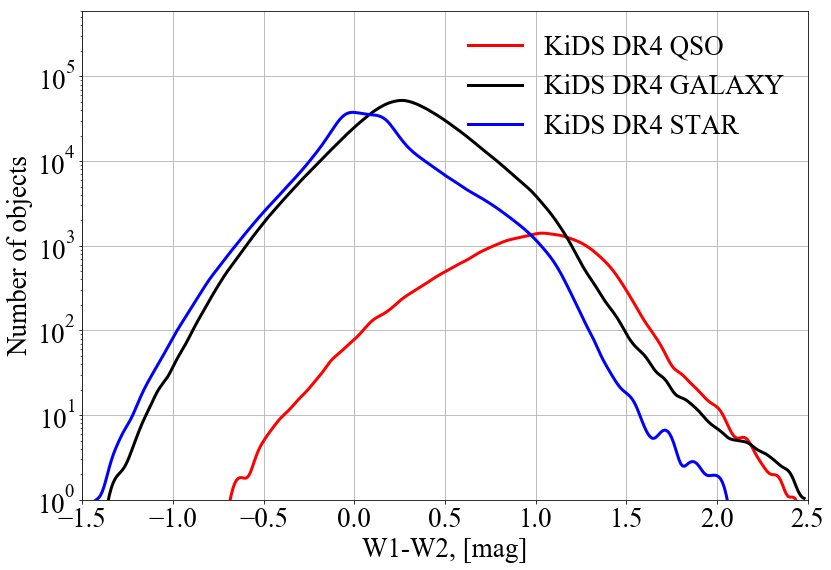}
\includegraphics[scale=0.3,angle=0]{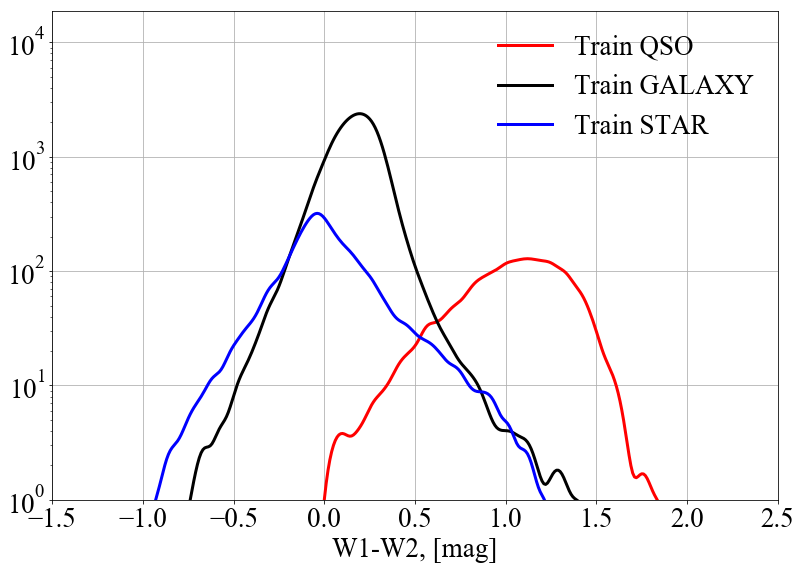}
\caption{Distribution of the $W1-W2$ colour for the classified KiDS DR4 sources of different classes (\textit{left} plot, \textit{solid} lines, \textit{red} for \texttt{QSO}, \textit{black} for \texttt{GALAXY}, \textit{blue} for \texttt{STAR}) versus their train samples (\textit{right} plot, \textit{dotted}, plotted with the same color-scheme). A fair agreement can be found between corresponding classes, although the distribution for the full catalogues is generally broader than the one of the training samples. The peak of the \texttt{QSO} distribution is shifted towards larger $W1-W2$ with respect to the \texttt{GALAXY} and \texttt{STARS}, as expected.}

\label{fig:allwise}
\end{figure*}

Figure~\ref{fig:allwise} shows the histograms of distribution of the $W1-W2$ color for the KIDS DR4 objects classified in the three classes (left panels, solid lines), and for the corresponding training sample (right panel, dotted lines), color coded by their classification: red for \texttt{QSO}, black for \texttt{GALAXY} and blue for \texttt{STARS}.  
In general, the distribution of the full catalogue shows a  
similar distribution to that of the training sample, with the peak of the \texttt{QSO} subsample shifted toward larger $W1-W2$ values, as expected. 
We note, however, that for the \texttt{GALAXY} and \texttt{STAR} classes, the distribution of the full catalogue is 
is much broader than the distribution of the corresponding training sample, especially towards larger values, both in negative and in  positive. 
This might indicate a lower pureness for these families and consequently a larger contamination level in the \texttt{QSO} family or a lack of particular class of families (e.g. active galaxies) in the training sample. 

As we show in the next subsection, the pureness of the objects classified as \texttt{GALAXY} seems to be quite high, according to the external validation of this class performed via a cross-match with the Galaxy And Mass Assembly Survey Data Release 3 \citep[GAMA DR3, ][]{GamaDr3}. 

More and deeper investigations will be performed on pureness and completeness in the forthcoming paper of the KiDS-SQuaD series. Nevertheless, for the purposes of this paper, we are confident, and we will prove in Section~\ref{sec:lensing}, that our automatic classifier allowed us to obtain a starting catalogue of quasars and galaxy with a stellar contamination much smaller than the one obtained in Paper I, where we rely on simple and manual optical and infrared color-cuts. 


\subsection{Validation of galaxies with GAMA}
To validate the pureness and completeness of the subsample of galaxies within the BEXGO catalogue, we cross-match the final catalogue of classified object with spectroscopically confirmed galaxies from GAMA.

In particular, following the suggestions given on the GAMA website, we retrieved all the objects\footnote{also the ones observed by other surveys, i.e. we queried the table 'SpecAll'} with redshifts $0.05<z<0.9$ and with a high "normalised" redshift quality ($nQ>1$). We matched these $\approx208$k sources with our final catalogue of classified objects from KiDS, finding 105\,334 systems in common. Among these 105\,018 were indeed classified as \texttt{GALAXY} from CatBoost and 104\,970 have a  $p_{\texttt{GALAXY}} \ge 0.8$. Thus, only the 0.3\% of the common objects have been misclassified (123 as \texttt{STARS} and 181 \texttt{QSO}).

Although we are aware that this test is not definitive and that it is not straightforward to directly translate the relative number of contaminant into a percentage of pureness of the final galaxy catalogue, it shows that, at least for this small but representative sample og galaxies, our CatBoost classifiers does a good job. 

We speculate that one of the reasons for slight disagreement between the distribution of galaxies from the training sample and galaxies classified as such in the BEXGO catalogue in the $W1-W2$ space might arise from the fact that, although we limited the analysis and classification to only objects brighter than $r<22^m$, the SDSS galaxies are generally more luminous than the KiDS ones.

We stress again that our final purpose is to create an authomatic and effective method to build a catalogue of extragalactic objects, with the smallest possible contamination from stars, that is the first necessary step to search for strong gravitational lensed quasars. We believe that these three validation steps with external data demonstrated that we succeeded in our goal and thus we can now use the newly created catalogue to search for lens candidates.

\section{Searching for gravitationally lensed quasars}
\label{sec:lensing}
Strong gravitationally lensed quasars are valuable but very rare objects (according to \citealt{Oguri10}, one quasars in $\sim 10^{3.5}$ is expected to be strongly lensed for $i$-band limiting magnitude deeper than $i\approx21^m$, see e.g. their Fig.~3 and Sec.~3.1) that give direct, purely gravitational probes of cosmology and extragalactic astrophysics. 

Generally speaking, we can separate lensed quasars in three families: systems where the quasar images dominate (mainly low-separation couples/quadruplets with a faint deflector in between), objects were the deflector is a bright, usually red, galaxy that dominates the light budget of the system, and finally systems where both lens and source give a non-negligible contribution to the light. In the last two cases, CatBoost will most probably return multiple matches of which at least one will be classified as extended (\texttt{GALAXY}), while in the former one it will classify them as \texttt{QSO}. However, we note that in cases where the separation between the quasar components is too low, the objects might be not resolved in the KiDS catalogue, and thus result in a single match/classification from our algorithm.  
Indeed, most of the known gravitationally lensed quasars with low separation between the multiple images, discovered in the SDSS, are identified as galaxies (only in few cases as single quasar), since the poor resolution does not allow deblending. Of course the better image resolution of KiDS helps in this case, however some lenses with very low separation are blended also in KiDS. 
This is why it is of crucial importance to have a catalogue of extragalactic objects as clean as possible from stellar contamination and as complete and efficient as possible in classifying galaxies and quasars. 
To demonstrate our statement that lensed quasars are not always classified as (multiples, near-by) \texttt{QSO} by ML based algorithms that work in a magnitude-color space, and at the same time to highlight the importance of having an extragalactic catalogue, we carried on a test on the recovery of known lenses, as already done in Paper I. 

\begin{table}
\center
\begin{tabular}{clll}
\hline
\hline
ID & $p_{\texttt{STAR}}$ & $p_{\texttt{QSO}}$ & $p_{\texttt{GALAXY}}$ \\  
\hline
\hline
& & & \\
J004941.90-275225.87 & 2.0E-5 & 1.0E-5 & \bf{0.9999}\\
& & & \\
J033238.22-275653.32 & 2.0E-5 & 1.0E-5 & \bf{0.9999}\\
& & & \\
J115252.26+004733.11 & 2.0E-5 & 1.0E-5 & \bf{0.9999}\\
& & & \\
J220132.76-320144.73 & 0.0004 & 4.0E-5 & \bf{0.9999}\\
& & & \\
J234416.95-305625.98 & 0.0004 & 0.0012 & \bf{0.9983}\\
& & & \\
J105644.89-005933.34 & 7.0E-4 & \bf{0.9978} & 0.0015\\
& & & \\
J112320.73+013747.53 & 0.0086 & \bf{0.9829} & 0.0085\\
& & & \\
J142758.89-012130.31 & 0.0035 & \bf{0.9831} & 0.0134\\
& & & \\
J025257.87-324908.65 & 0.0011 & \bf{0.9950} & 0.0039\\
\hline
& & & \\
J145847.59-020205.87 & 0.0004 & 0.0011 & \bf{0.9985}\\
J145847.66-020204.86 & 2.0E-5 & 3.0E-5 & \bf{0.9999}\\
& & & \\
J032606.87-312254.21 & 0.0019 & \bf{0.9944} & 0.0037\\
J032606.78-312253.52 & 0.0023 & \bf{0.9793} & 0.0184\\
& & & \\
J143228.96-010613.51 & 0.0006 & \bf{0.9980} & 0.0014\\
J143229.25-010615.98 & 0.0008 & \bf{0.9966} & 0.0025\\
& & & \\
J104237.27+002301.42 & 0.0477 & \bf{0.8637} & 0.0886\\
J104237.24+002302.76 & 0.0652 & \bf{0.7721} & 0.1627\\
\hline
& & & \\
J092455.82+021923.69 & 0.0078 & \bf{0.8909} & 0.1012\\
J092455.82+021925.30 & 0.0059 & 0.4543 & \bf{0.5397}\\
& & & \\
J122608.10-000602.31 & 0.0046 & \bf{0.9610} & 0.0343\\
J122608.03-000602.25 & 0.0199 & \bf{0.5048} & 0.4752\\
J122608.13-000559.09 & 0.0009 & 0.0016 & \bf{0.9997}\\
& & & \\
J133534.80+011805.61 & 0.0128 & \bf{0.9806} & 0.0066\\
J133534.87+011804.45 & 0.0056 & \bf{0.9797} & 0.0066\\
J133534.97+011809.32 & 0.0013 & 0.0050 & \bf{0.9937}\\
& & & \\
J152720.14+014139.66 & 0.0058 & \bf{0.9617} & 0.0325\\
J152720.27+014140.96 & 0.0005 & 0.0006 & \bf{0.9999}\\
\hline
\end{tabular}
\caption{Known lenses in the KIDS DR4 footprints. All of them are recovered in our extragalactic catalogue, 8 have multiple matches. For the single ones (upper 'block'), 5 are classified as \texttt{GALAXY} and 4 as \texttt{QSO} (we highlight in bold the highest probability). For the multiple matches, half of the time all the components belong to the same family (middle 'block') and the other half, they belong to different families. We report for each component of each system the J2000 coordinates in the ID column (in the usual 'hhmmss.ss$\pm$ddmmss.ss' format), and the probability to belong to each of the three classification families.}
\label{tab:known_lenses}
\end{table}

We started from the same list of $\approx260$ confirmed lensed quasars that we used in Paper I, collected from the CfA-Arizona Space Telescope LEns Survey \citep[CASTLES, ][]{Munoz98} Project database, the SDSS Quasar Lens Search (SQLS, \citealt{Inada12}) and updated it with systems recently discovered in wide-sky surveys \citep{Agnello18_atlas, Agnello18_gaia, Ostrovski18, Anguita18, Lemon18, Spiniello19_des}.  
The cross-match between the full KIDS DR4 catalogue of objects with $r<22^m$ and the list of known lensed quasars (288 systems), gave us 17 known lenses. All of them have been retrieved in the KiDS-BEXGO catalogue, 10 classified as \texttt{QSO} and 7 as \texttt{GALAXY} (one of them with a $p_{\texttt{QSO}}\approx0.45$). These lenses are reported in Table~\ref{tab:known_lenses}, together with the probability to belong to each class. We do not explicitly report their right ascensions and declinations in separate columns because their ID already contain the J2000 coordinates in the usual 'hhmmss.ss$\pm$ddmmss.ss' format.

Based on this simple and qualitative test, it appears clear that selecting only quasars would allow one to find only lens systems were the contribution to the light from the quasars is much larger than the contribution of the deflector (selecting only \texttt{QSO} we would retrieve roughly 65\% of the known lensed quasar population -- 11 over 17 systems).

Finally, although this goes behind the scope of Paper II, we note that another important advantage of having an extragalactic source catalogue (rather than only quasars) is the possibility to search for galaxy-galaxy gravitational lenses 

Such type of gravitationally lensed objects allow to investigate in great details the mass distribution in massive galaxies upt to $z\sim1$, especially when combined with dynamics \citep{Koopmans06, Koopmans09,Spiniello11, Spiniello15}. Morphological and photometric criteria can be used to find this kind of lenses: one should look for red extended objects (\texttt{GALAXY} with red colors), with the presence of blue extended objects (\texttt{GALAXY} with blue colors) within small circular apertures. We will work in this direction in a forthcoming paper, possibly using authomatic, machine learning based routines to this scope (e.g. \citealt{Petrillo17, Petrillo19a, Petrillo19b}) and already available catalogues of Luminous red galaxies in the Kilo Degree Survey (e.g., \citealt{Vakili2019}. 


Starting from the KiDS-BEXGO catalogue of 5\,880\,276 objects, we retrieve only systems belonging to the following distinct groups:
\begin{enumerate}
\item QSO-Multiplets: sources classified as \texttt{QSO} and with at least one near-by \texttt{QSO} companion (within a $5\arcsec$ circular aperture radius) with similar colors,
\item GALAXY-Multiplets: sources classified as \texttt{GALAXY} and surrounded by at least one object classified as \texttt{QSO} within a $5\arcsec$ circular aperture radius\footnote{The choice of a $5\arcsec$ circular aperture radius is motivated by the average separation of all the known lenses.}
\end{enumerate}

\begin{figure*}
\center
\includegraphics[scale=0.28,angle=0]{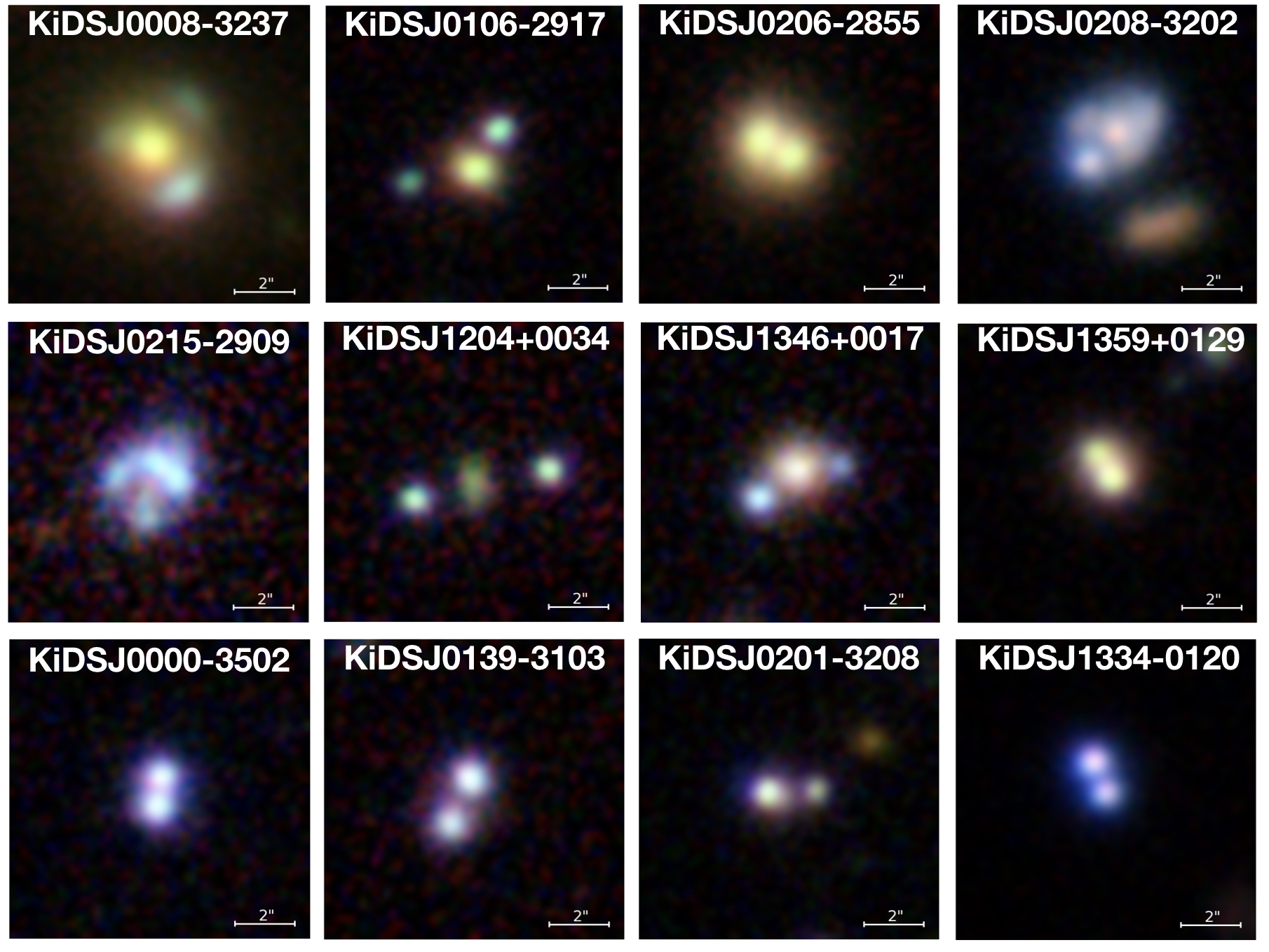}
\caption{The 12 best candidates, both of GAL-Multiplets type (first two rows) and of QSO-Multiplets type (bottom row). The cutouts show $gri$ (for the moment ugr) combined KiDS images of $10\arcsec\times10\arcsec$ in size. The coordinates of the candidates are given in Table~\ref{tab:qso_candidates}.}
\label{fig:candidates}
\end{figure*}

This simple procedure allowed us to obtain 347 unique objects for the first group and 611 unique objects belonging to the second one, which we then visually inspected. 
Among these, some where already known lenses, some are probably binary quasars and some are simply contaminants appearing asclose-by companions because of sky projection. 
Nevertheless, we found many very promising lens candidates, that two people of our team graded from 0 to 4, with 4 being a sure lens. 
We present the 12 candidates with grade $\ge2.5$ in Table~\ref{tab:qso_candidates} (divided into the two Multiplets kinds). We publicly release their coordinates to facilitate spectroscopic follow-up, which is the last necessary step for the unambiguous confirmation. Finally, the $gri$-combined KiDS cutouts of these 12 candidates are shown in Figure.~\ref{fig:candidates}. The first top rows show candidates belonging to the  GALAXY-Multiplets family while the bottom row show QSO-Multiplets candidates. In the former group, the deflector give a much larger contribution to the light, as can be seen from the images. KiDSJ0008-3237 seem to be a very reliable galaxy-galaxy candidate, while KiDSJ0215-2909, definitively among the most promising objects, might be a fold-quadruplet, similar to the one recently found in the VST-ATLAS Survey, WISE 025942.9-163543 \citep{Schechter18}.
and very useful for cosmography studies (time-delay measurement of $H_0$, see e.g.'The H0 Lenses in COSMOGRAIL's Wellspring'\footnote{\url{https://shsuyu.github.io/H0LiCOW/site/}} results).

We note that, of the 17 known lenses, only 8 have been selected as multiplets (2 as QSO-Multiplets and 6 as GALAXY-Multiplets). The other 9 systems have not been deblended in the KiDS catalogue, and thus only have one single match in our classification scheme.   
These numbers are perfectly in line with the results obtained in Paper I where we found that the Multiplet method alone allowed the recovery of $\sim40$\% of the known lenses population. In a forthcoming paper of this series, fully dedicated to the lens search, we will perform a more careful candidate selection, also based on improved color and magnitude criteria to select objects with similar colors and applying to the full BEXGO catalogue the Blue and Red Offsets of Quasars and Extragalactic Sources (BaROQuES) scripts, already successful tested in Paper I.

We finally note that we re-discovered a very promising quadruplet: KIDS0239-3211 that was presented in a AAS research note \citep{Sergeyev2018}  and it was found by the first application of our ML based classifier\footnote{We used in that case a Random Forests classifier, trained with spectroscopically confirmed objects from SDSS DR14.}. The same system has also been detected \citealt{Hartley17} using image-based Support Vector Machine classifier and by \citealt{Petrillo19b} using Convolutional Neural Networks; but since they do not released the coordinates in their paper, we re-discover it in a completely independent way. 

\begin{table*}
\center
\begin{tabular}{ccccccc}
\hline
\hline
 & & & & & & \\
ID & RA & DEC & Multiplet  & Num. of    & Separation  & Notes\\
 & (hh:mm:ss.ss) & ($\pm$dd:mm:ss.ss)   & type       & matches  & (arcsec) & \\
\hline
KIDSJ0008-3237 & 00:08:16.01 & -32:37:15.80 & GAL & 3  & 2.7 & Gravitational arc\\
KIDSJ0106-2917 & 01:06:49.80 & -29:17:12.20 & GAL & 2  & 3.4 & Double with large shear\\
KIDSJ0206-2855 & 02:06:30.86 & -28:55:42.22 & GAL & 2  & 1.2 & Low-separation double\\
KiDSJ0208-3203 & 02:08:53.16 & -32:02:03.51 & GAL & 3	& 2.0 & Cross Quad candidate \\
KIDSJ0215-2909 & 02:15:14.4 & -29:09:25.6 & GAL & 3 &  3.2 & Fold Quad candidate\\ 
KIDSJ1204+0034 & 12:04:56.58 & +00:34:06.02 & GAL & 3 & 4.6 & Large-separation double\\
KiDSJ1346+0017 &  13:46:12.38 & +00:17:20.18 & GAL & 4 & 2.8 & Double with large shear\\
KIDSJ1359+0129 & 13:59:43.98 & +01:28:13.90 & GAL & 2  & 0.9 &  Low-separation double\\
\hline
KIDSJ0000-3502 & 00:00:57.10 & -35:02:54.15 & QSO & 2  &  1.0 &  Low-separation double\\
KIDSJ0139-3103 & 01:39:59.08 & -31:03:35.06 & QSO & 2  & 1.6 &  Low-separation double\\
KIDSJ0201-3208 & 02:01:15.44 & -32:08:34.57 & QSO & 2  & 1.6 &  Low-separation double\\
KIDSJ1334-0120 & 13:34:11.18 & -01:20:52.22 & QSO & 2  & 1.1 &  Low-separation double\\
\hline
\end{tabular}
\caption{The most reliable gravitationally lensed quasar candidates found with the multiplet method described in the text. Cutouts of the candidates are shown in Figure~\ref{fig:candidates}. We highlight in the table the number of matches found for each system in the BEXGO catalogue. In groups where the number of detection is smaller than the total number of objects visible from the cutout, most probably these objects are fainter than the magnitude threshold we set for the input catalog. We double checked that these missing matches were not objects classified as stars. The last column indicates the separation between the multiple QSO images.}
\label{tab:qso_candidates}
\end{table*}

Finally, we cross-match the list of all the lens candidates found in Paper I with the BEXGO catalogue. 
We find that among the 210 objects we have found in \citealt{Spiniello18}, 148 are recovered in the extragalactic objects catalogue ($\approx45\%$ classified as \texttt{QSO} and $\approx55\%$ as \texttt{GALAXY}) and 66 are also selected as Multiplets. Of the 62 remaining objects, 33 have $r>22^m$ and therefore were discarded at the input catalogue creation stage, and 29 were classified as \texttt{STAR} by CatBoost; these 29 stars indeed also have a match in \textit{Gaia}, and all of them have non-negligible proper motions and parallaxes.
Finally, among the DR3 candidates that have not been found in the DR4 KiDS-BEXGO, 4 have been spectroscopically followed-up and turned out to be stars\footnote{We have already started a spectroscopic follow-up campaign using different facilities (e.g. the NTT@La Silla, the SALT@Suthernland Observatory, the LBT@Mt.~Graham). The detailed results will be presented in forthcoming dedicated papers.}. 
These numbers nicely demonstrate that the employment of a ML based classifier further help in decreasing the risk of stellar contamination within gravitationally lensed quasar candidates. 

Of the seven known lenses that we recovered in Paper I, six are still recovered. We only lose the nearly Identical Quasar (NIQ) couple QJ0240-343 \citep{Tinney95, Tinney97} behind the Fornax dwarf spheroidal galaxy, once again because it has $r$ mag of $r=22.17^m$ and thus it does not satisfied our initial conditions. 

\section{Results and Conclusions}
\label{sec:conclusions}
In this second paper of the KiDS Strongly lensed QUAsar Detection Project (KiDS-SQuaD) we have presented a new machine learning based classifier to identify extragalactic objects in order to find lensed quasars within the KiDS DR4 data. 
The technique adopted in this paper became quite standard in the extragalactic community
to classify objects in multi-band photometric surveys \citep{Gieseke2011,ksz2012,Brescia2015,Carrasco15,Peters15,Krakowski2016,Krakowski2018,Viquar2018,Khramtsov2018,Barrientos2018,Nolte2019,Bai2019}, which provide a very large amount of data, and have been already tested on the KiDS DR3 \citep{Nakoneczny2019}. 

In fact, \citet{Nakoneczny2019} presented a ML based pipeline that allowed them to classify objects into three classes (stars, galaxies and quasars) and successfully applied it to the KiDS DR3. Our work, although extending from their findings, has been developed within a different framework, i.e. the search of lensed quasars, and it therefore differs from \citetalias{Nakoneczny2019} in many aspects, from the assumption that quasars are point-like source, to the cleaning procedure, optimization, and fine-tuning aimed at minimize as much as possible the stellar contamination in the catalogue of extragalactic objects. Finally, here we also add infrared data, using deep photometry in 9-bands (instead of four), which further helps in isolating stars.

\subsection{Summary}
\label{sec:summary}
We provide here a general summary of the archived results of this paper,  highlighting with bullet points the main steps that we undertook from the presentation of a new pipeline to the search of gravitationally lensed quasars. In particular, we have:
\begin{itemize}
\item used the full potential of machine learning methods on broad optical-infrared photometry data, after having applied a careful cleaning on the training SDSS$\times$KiDS sample, also visually inspecting the ambiguous cases when necessary; 
\item performed an ad-hoc customization and fine-tuning of the parameters of the CatBoost algorithm, which we identified as the best possible classifiers for our purposes, to reach the required levels for purity and completeness and to avoid overfitting poroblems. We also implemented a weighting procedure, that allowed us to reach the best possible purity of quasars (decreasing the rate of stars, classified as quasars, from 0.6\% to 0.3\%); 
\item splitted the training dataset into a hold-out and out-of-fold part to asses the performance of our classifier in terms of completeness and pureness; 
\item defined (and then solved) a 3-class problem (\texttt{STAR}, \texttt{GALAXY}, \texttt{QSO}), working with a simple basic assumption made for the classification, namely that quasars and stars are point-like sources while galaxies are extended. We therefore used the \texttt{CLASS\_STAR} parameter -- a 'stellarity' index from KiDS catalogue --  which turned out to be the most important feature in our classification algorithm (as in \citetalias{Nakoneczny2019}), together with optical and infrared colors; 
\item applied CatBoost on all the data from KiDS DR4 with magnitude brighter than $r=22^m$. For each source, the classifier calculated the probability of belonging to the three different classes of objects: $p_{\texttt{STAR}}$, $p_{\texttt{GALAXY}}$, $p_{\texttt{QSO}}$, and then we assumed that a source belongs to a given class when the probability of being in that class is the highest; 
 \item studied the variation in completeness and pureness as a function of the probability threshold used to assign an object to a given class;
\item collected all the objects that were not classified as stars, building the KiDS DR4 Bright EXtraGalactic Objects catalogue (KiDS-BEXGO), which we then also validated using external data (\textit{Gaia} DR2, AllWISE and GAMA);  
\item showed the potential of the KiDS-BEXGO catalogue in the gravitationally lensed quasar search, with a simple test on the recovery of known, confirmed lenses, and proved, in this way, that our method of selecting extragalactic sources (not only quasars) is a necessary condition to discover as many new systems as possible; 
\item used the KiDS-BEXGO catalogue to search for new, undiscovered gravitationally lensed quasars, looking for objects with a near-by companion. We have obtained a list of 958 'Multiplets' (347 \texttt{QSO} and 611 \texttt{GALAXY}) that we visually inspected, finding 12 very reliable lens candidates for which we release coordinates and KiDS images; 
\item showed the improvement, in terms of stellar contaminants in the final candidate list with respect to what obtained in in Paper I, but at the same time also demonstrated the need for different methods to search for lenses candidates within the catalogue (e.g. the BaROQuES) and directly analyzing images (DIA). These methods will be investigate in a forthcoming publication. 
\end{itemize}

In addition, we present in the Appendix~\ref{testing_classifiers} a direct comparison of some of the most used classifiers based on decision trees.  
This test helped us to compare and quantify the performance of each of them on the same training sample in order to choose the most suitable one for our purposes, namely CatBoost. 

\subsection{Future perspectives and improvements}
\label{sec:future}
From the predictions of \citet{Oguri10}, we estimated that $\approx50$ lensed quasars are expected in the KIDS DR4 (1000 deg$^2$), when limiting to systems with $r<22^m$; 17 lenses are already known, thus, in principle, more than half are still undiscovered (and even more going to fainter magnitudes). 

In this Paper II, we focused on the first necessary step to find all the catchable gravitational lenses: an object classifier, that allowed us to get rid of the very numerous stellar contaminants and will allow us to analyze with a minimum human intervention very large datasets. 
We note that our classifier is build and trained for this specific purpose. 
A forthcoming paper within the KiDS consortium (Nakoneczny et al., in prep.)
will present a machine learning based pipeline trained for general scientific purposes, providing photometric redshifts for galaxies and quasars in KiDS DR4 on top of the objects classification, and testing machine learning extrapolation to increase catalog completeness on fainter magnitudes.

 
Moreover, we also plan to further improve the classification model, working in a more complex and complete feature space and developping a more detailed classification scheme (e.g., spitting the classification of galaxies on late and early types, giving that massive early types are more likely acting as deflectors because on average more massive). 

In Paper III, already in preparation (Sergeyev et al., in prep), we focus instead only on the gravitationally lens search, presenting a more systematic way, as automatic as possible, to select reliable candidates from the KiDS-BEXGO catalogue. We will apply photometric and morphological criteria, e.g. based on optical and infrared color, or on the simple fact that a centroid offsets of the same object among different surveys, covering different bands is expected since the deflector and quasar images contribute differently in different wavelength ranges (BaROQuES). We will also exploit the full potential of the Direct Image Analysis (DIA, see Paper I for more details) to get precise astrometry and fit the photometry of our most reliable candidates. 

Finally, we already started the necessary spectroscopic follow-up, to get a final, unambiguous confirmation of the lensing nature for as many systems as possible, and to obtain secure redshift measurements that will allow us translate the lens model results (e.g., Einstein radii) into physical mass measurements.

\vspace{1.6cm}

\begin{acknowledgements}
The authors wish to thank Maciej Bilicki for the interesting discussion and the very constructive comments that helped in improving the final manuscript.  CS has received funding from the European Union's Horizon 2020 research and innovation programme under the Marie Sk\l odowska-Curie actions grant agreement No 664931. 
CT acknowledges funding from the INAF PRIN-SKA 2017 program 1.05.01.88.04.
KK acknowledges support by the Alexander von Humboldt Foundation.
JTAdJ is supported by the Netherlands Organisation for Scientific Research (NWO) through grant 621.016.402.
HYS acknowledges the support from the Shanghai Committee of Science and Technology grant No. 19ZR1466600.

This work is based on data products from observations made with ESO Telescopes at the La Silla or Paranal Observatories under programme ID(s) 177.A-3016(A), 177.A-3016(B), 177.A-3016(C), 177.A-3016(D), 177.A-3016(E), 177.A-3016(F), 177.A-3016(G), 177.A-3016(H), 177.A-3016(I), 177.A-3016(J), 177.A-3016(K), 177.A-3016(L), 177.A-3016(M), 177.A-3016(N), 177.A-3016(O), 177.A-3016(P), 177.A-3016(Q), 177.A-3016(S), 177.A-3017(A), 177.A-3018(A), 060.A-9038(A), 094.B-0512(A), and on data products produced by Target/OmegaCEN, INAF-OACN, INAF-OAPD and the KiDS production team, on behalf of the KiDS consortium. 
OmegaCEN and the KiDS production team acknowledge support by NOVA and NWO-M grants. Members of INAF-OAPD and INAF-OACN also acknowledge the support from the Department of Physics \& Astronomy of the University of Padova, and of the Department of Physics of Univ. Federico II (Naples).
This publication makes use of data products from the Wide-field Infrared Survey Explorer, which is a joint project of the University of California, Los Angeles, and the Jet Propulsion Laboratory/California Institute of Technology, and NEOWISE, which is a project of the Jet Propulsion Laboratory/California Institute of Technology. 
WISE and NEOWISE are funded by the National Aeronautics and Space Administration.

This publication makes use of data products from the Sloan Digital Sky Survey IV. Funding for SDSS IV has been provided by the Alfred P. Sloan Foundation, the U.S. Department of Energy Office of Science, and the Participating Institutions. 
SDSS-IV acknowledges support and resources from the Center for High-Performance Computing at the University of Utah. The SDSS web site is www.sdss.org.
SDSS-IV is managed by the Astrophysical Research Consortium for the Participating Institutions of the SDSS Collaboration including the Brazilian Participation Group, the Carnegie Institution for Science, Carnegie Mellon University, the Chilean Participation Group, the French Participation Group, Harvard-Smithsonian Center for Astrophysics, 
Instituto de Astrof\'isica de Canarias, The Johns Hopkins University, 
Kavli Institute for the Physics and Mathematics of the Universe (IPMU) / 
University of Tokyo, Lawrence Berkeley National Laboratory, 
Leibniz Institut f\"ur Astrophysik Potsdam (AIP), 
Max-Planck-Institut f\"ur Astronomie (MPIA Heidelberg),
Max-Planck-Institut f\"ur Astrophysik (MPA Garching), 
Max-Planck-Institut f\"ur Extraterrestrische Physik (MPE), 
National Astronomical Observatories of China, New Mexico State University, 
New York University, University of Notre Dame, 
Observat\'ario Nacional / MCTI, The Ohio State University, 
Pennsylvania State University, Shanghai Astronomical Observatory, 
United Kingdom Participation Group,
Universidad Nacional Aut\'onoma de M\'exico, University of Arizona, 
University of Colorado Boulder, University of Oxford, University of Portsmouth, 
University of Utah, University of Virginia, University of Washington, University of Wisconsin, Vanderbilt University, and Yale University.

 This work has made use of data from the European Space Agency (ESA) mission
 \textit{Gaia} (\url{https://www.cosmos.esa.int/gaia}), processed by the \textit{Gaia}
 Data Processing and Analysis Consortium (DPAC,
 \url{https://www.cosmos.esa.int/web/gaia/dpac/consortium}). Funding for the DPAC
 has been provided by national institutions, in particular the institutions
 participating in the \textit{Gaia} Multilateral Agreement.

This work has made use of data from Galaxy and MAss Assemby Survey. GAMA is a joint European-Australasian project based around a spectroscopic campaign using the Anglo-Australian Telescope. The GAMA input catalogue is based on data taken from the Sloan Digital Sky Survey and the UKIRT Infrared Deep Sky Survey. Complementary imaging of the GAMA regions is being obtained by a number of independent survey programmes including GALEX MIS, VST KiDS, VISTA VIKING, WISE, Herschel-ATLAS, GMRT and ASKAP providing UV to radio coverage. GAMA is funded by the STFC (UK), the ARC (Australia), the AAO, and the participating institutions. The GAMA website is http://www.gama-survey.org/.

\end{acknowledgements}

\begin{appendix}
\section{Testing different classifiers}\label{testing_classifiers}
The main result of the main body of the paper, 
is a catalogue of bright objects from KiDS DR4 
that we classified into three families, 
\texttt{STAR}, \texttt{QSO},\texttt{GALAXY}, using a 
machine learning based classifier, that uses the Gradient Boosting algorithm. 

In order to choose the best possible algorithm for our purposes, we tested two different approaches (and three diffrent methods) based on ensembles of decision trees, namely the Gradient Boosting (GB, \citealt{gradboost}) and Random Forest \citep[RF, ][]{RF} algorithms, which was the choice of \citetalias{Nakoneczny2019}. 
Here in this Appendix, we provide a more detailed description of the classifiers, their main characteristics, strength and weakness points to give to the reader a better understanding of the differences and similarities between them and to finally justify our final choice. 

As already stated in the main body, the general classification problem can be simply explained considering a training dataset ($D$) with $n$ samples and $m$ features for each sample with defined label $y_i$: $D=\{\mathbf{x}_i,y_i\}$ where $i\in\{0,...,n-1\}, \mathbf{x}_i \in \mathbb{R}^m, y_i \in \mathbb{N}$. The goal is then to create an approximation function $F: \mathbf{x}\to y$. 

The two GB algorithms mentioned above follow two different approaches of ensemble learning, that could be propagated not only on the decision trees. We describe them in details in the following sections.


\begin{figure*}[t]
\center
\includegraphics[scale=0.21,angle=0]{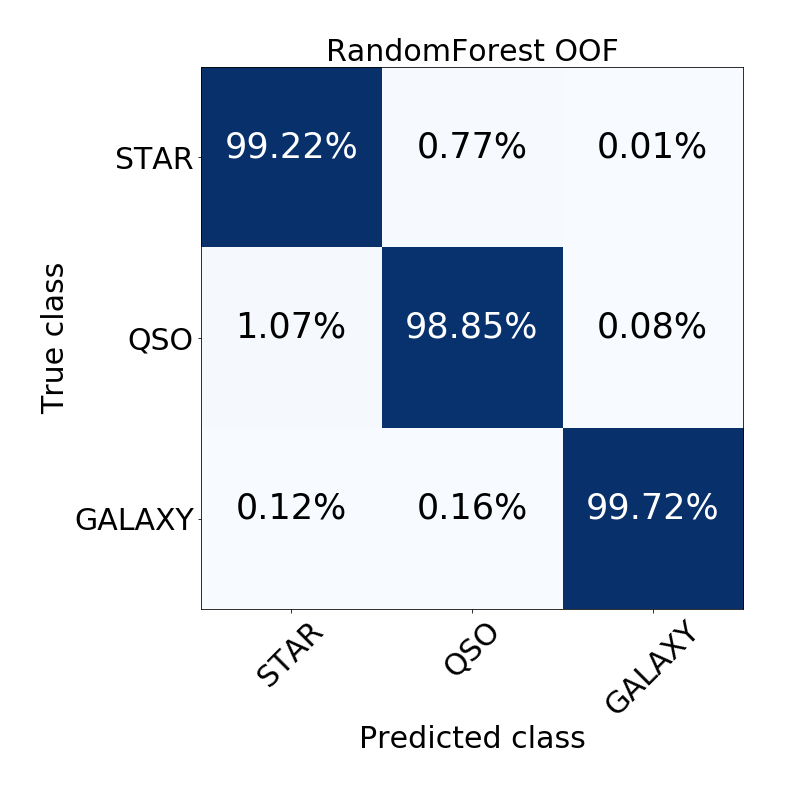}
\includegraphics[scale=0.21,angle=0]{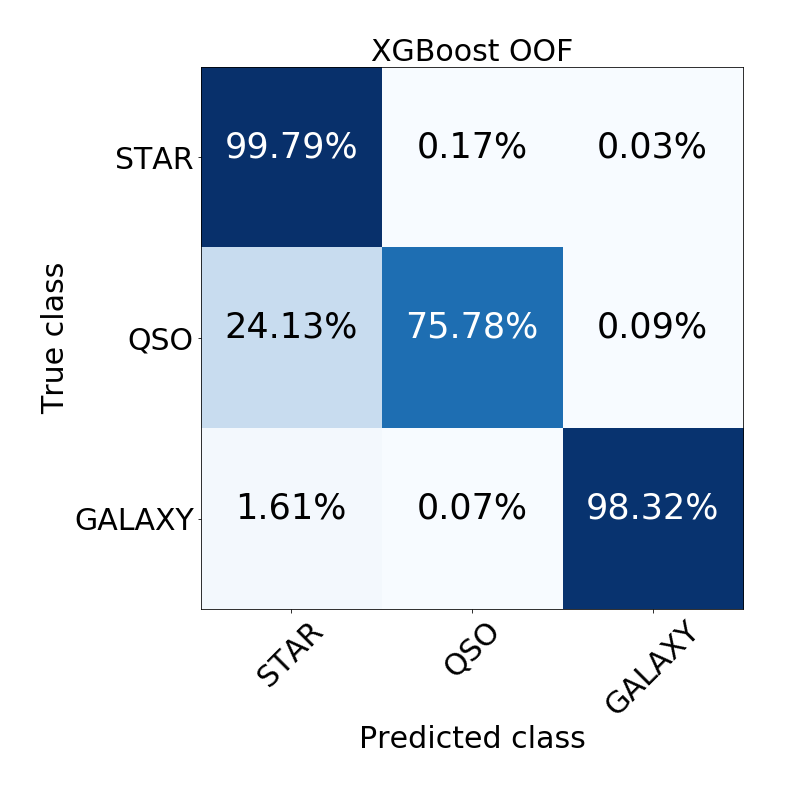}
\includegraphics[scale=0.21,angle=0]{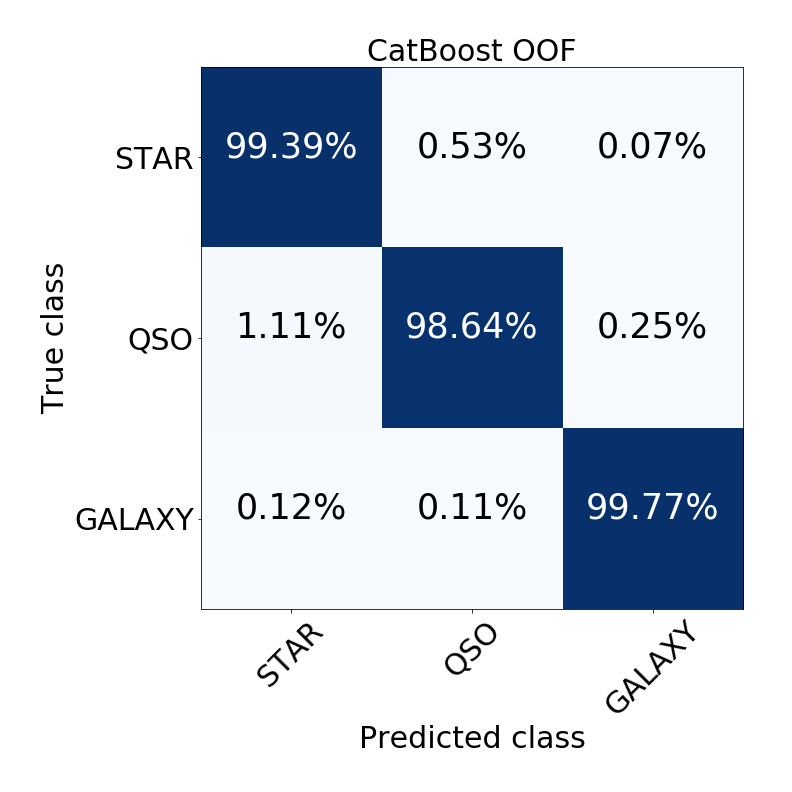}
\includegraphics[scale=0.21,angle=0]{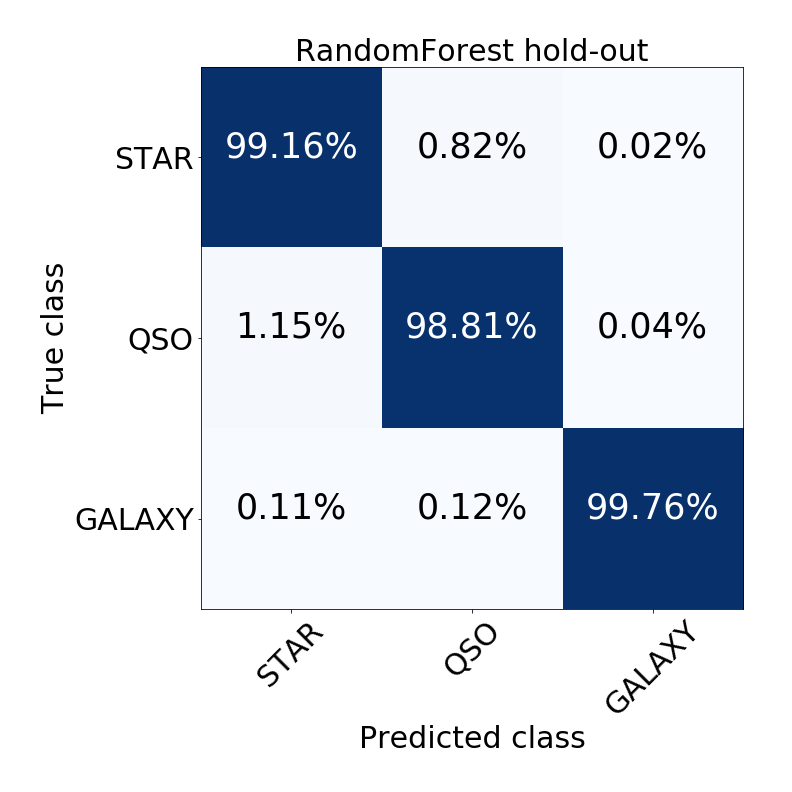}
\includegraphics[scale=0.21,angle=0]{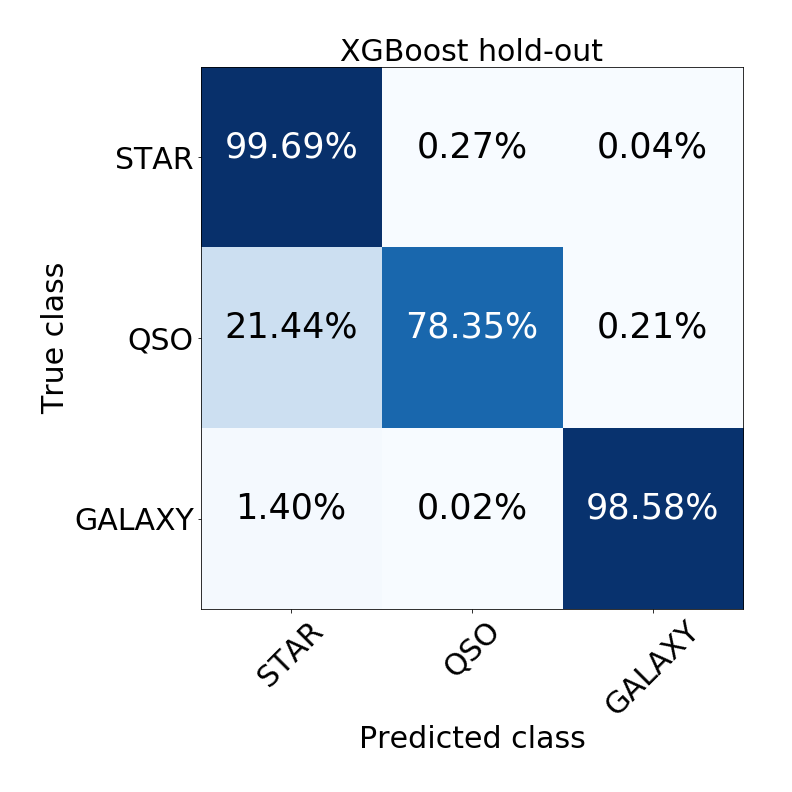}
\includegraphics[scale=0.21,angle=0]{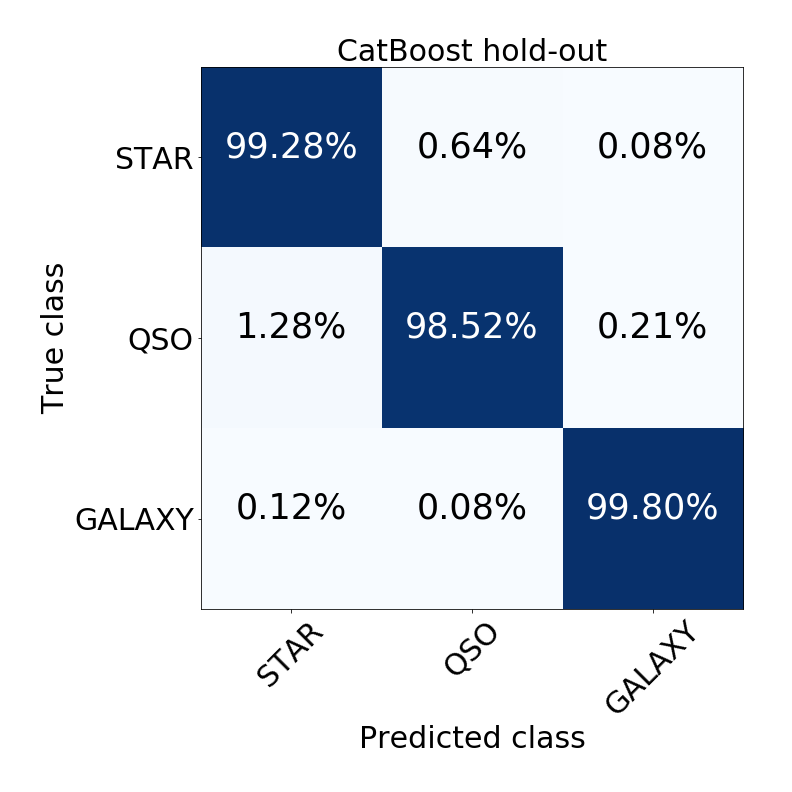}
\caption{Confusion matrices for the RF (\textit{left}), XGBoost (\textit{central}) and CatBoost (\textit{right}) predictions on OOF sample (\textit{top} panels) and hold-out sample (\textit{bottom} panels).}
\label{fig:confmatr}
\end{figure*}

\subsection{Gradient Boosting Decision Trees}
\label{GBDT}
XGBoost, \citealt{xgb} and CatBoost, \citealt{cat1,cat2}), 
are two Gradient Boosting \citep[GB, ][]{gradboost} algorithms that implement two different schemes for calculating gradients. 

Let us consider the ensemble of $K$ trees, where the predicted score for an input $\mathbf{x}$ is given by the sum of the values predicted by the individual $K$ trees: $\hat{y}^K(\mathbf{x})=\sum_{j=1}^K f_j(\mathbf{x})$, where $f_j$ is the output of the $j$-th decision tree. 

Building the $(K+1)$th decision tree minimizes an objective function $L=\sum_{i=1}^n \ell\big(y_i, \hat{y_i}^K(\mathbf{x}_i) + f_{K+1}(\mathbf{x}_i)\big)+\Lambda(f_{K+1})$, where $\ell\big(y_i, \hat{y_i}^K(\mathbf{x}_i) + f_{K+1}(\mathbf{x}_i)\big)$ depends on the first (and, possibly, second) deviation of the loss function $\ell\big(y_i, \hat{y_i}^K(\mathbf{x}_i)\big)$, and $\Lambda(f_{K+1})$ is a regularization function that penalizes the complexity of the $(K+1)$-th tree to prevent overfitting.
To build a $(K+1)$-th decision tree, the algorithm starts with a single decision node and iteratively tries to add a best split for each node, until a stop criterion on tree growth is satisfied. 

XGBoost estimates the gradient value for all of the objects in a leaf and calculates the average gradient to determine the best split for each node. In this way, the gradient is estimated via the same data points, on which the current decision tree was built on. 
In general, such splitting procedure leads to the gradient bias (due to the repeated usage of the same objects through all iterations) and, as result, to an overfitting problem \citep{cat2}. 

CatBoost, the chosen algorithm for this paper, in its turn, implements the splitting technique called Ordered Boosting \citep{cat2}, that overcomes this problem. With Ordered Boosting, the gradients are calculated not for all of the objects, but for the shuffled training dataset (so-called, random permutations), wherein the gradients are calculated for the objects before a given one. In such a way, gradient for the $j+1$ object is calculated based on prediction of the model, learnt by previous samples in shuffled dataset.


One of the limitations of the GBDT algorithms is a wide range of parameters that have to be tuned to get the highest classification quality; CatBoost has advantages also on this, because it performs well also without hyperparameters tuning.

For our task, the most influential hyperparameters, that have to be tuned in GBDT algorithms, are:
\begin{enumerate}
\item[1.] \texttt{learning\_rate} -- the rate of gradient descent;
\item[2.] \texttt{min\_split\_loss} -- the minimum loss reduction required to split a node of the tree;
\item[3.] \texttt{max\_depth} -- the maximum depth of the decision trees;
\item[4.] \texttt{min\_child\_weight} -- the minimum number of samples in the node of the decision tree required to make a split;
\item[5.] \texttt{max\_delta\_step} -- the maximum step controlling convergence during gradient descent;
\item[6.] \texttt{colsample\_bytree} -- the subsample ratio of the features during building each decision tree;
\item[7.] \texttt{subsample} -- the subsample ratio of the training objects
\end{enumerate}

In particular, we noticed that the parameters that mostly affected the learning quality for our training dataset were \texttt{learning\_rate} (greater values correspond to a sharper gradient descent, that is good for learning acceleration, but can lead to missing the loss minimum), and  \texttt{max\_depth} (greater values correspond to a large complexity of the trees, and can lead to overfitting).

Moreover, GB algorithms usually allow to use a stop criterion, responsible for the termination of the learning when an overfitting occurs (the so-called, \texttt{early\_stopping} parameter). It is expressed via the number of constructed trees, after which the quality of the metric does not increase anymore. Usually this parameter ranges between 10 and 1000 trees, depending on \texttt{learning\_rate}. If the early stopping criterion is met, the GB algorithm accepts the number of trees, satisfied to the best score.
Paying particular attention to the \texttt{early\_stopping} parameter is the best way to avoid as much as possible overfitting.
To quantify, how the quality of the classification changes over the iterations, for XGBoost and CatBoost, it is necessary to define a quality metric. Defining this metric allows to express the changing of the classification quality against the complexity of GBDT algorithm and can be easily used to control the overfitting. 
Widely used quality metrics are accuracy, precision, recall and F1-score; however, these metrics are sensitive to the imbalance in number of training sources among different classes. Therefore, we decided to use Matthews correlation coefficient \citep[MCC,][]{MCC} which is instead insensitive to it.

Finally, a good way to decrease the number of the stars and galaxies classified by the algorithm as quasars is to apply a weight to the loss function of these two classes. 
In fact, this trick, applied to CatBoost, helped us in the paper to improve the final purity of the quasar selection with a minimal decreasing of the completeness on the training set. In particular, we weighted the loss function for the \texttt{STAR} and \texttt{GALAXY} samples in the following way:
\begin{equation}
    L=\frac{1}{\sum_{i=1}^n w_i} \sum_{i=1}^n w_i [\ell\big(y_i, \hat{y_i}^K(\mathbf{x}_i) + f_{K+1}(\mathbf{x}_i)\big)+\Lambda(f_{K+1})]
\end{equation}

where $w_i = 1$ if source is a \texttt{QSO} and $w_i=4$ if source is \texttt{STAR} or \texttt{GALAXY}.


\subsection{Random Forest}
Another method of ensemble learning with decision trees is based on the use of a Random forest \citep[RF, ][]{RF} algorithm. This was the choice adopted in KiDS DR3. The basic idea of RF is that a set of decision trees can fit a robust classifier by averaging their decisions. 
From the performance side, the RF constructs a large number of decision trees and then uses the majority vote among them. 

The main pipeline of the single decision tree carries on the following steps:
\begin{enumerate}
\item generation of different samples from the training set with the same size, but using random subsets of all the objects. The repetition of some objects among different subsamples is required to make the sample complete (so-called, random subsample with replacement);
\item let the decision tree learning on the random subsample with replacement (1), but using randomly selected $\approx \sqrt{m}$ features. 
\end{enumerate}
The RF consists, then, in many learning processes, each performed on a single decision tree using different random subsamples with replacement and randomly selected features.
Then, the single prediction of a class for given objects is a simple average on the predictions of all constructed decision trees (bootstrap bagging method). 

The big advantage of RF is that it uses together the bootstrap bagging method (averaging prediction of the estimators learnt with random subsamples with replacement) and the learning of each estimator with random subset of features. 
These procedures prevent overfitting and, in most of the cases, help to improve the  classification performance and to increase the generalization ability of the RF.

The principal hyperparameters, that are required to the RF fitting on the training dataset, are:
\begin{enumerate}
\item[1.] \texttt{n\_estimators} -- the number of decision trees;
\item[2.] \texttt{max\_features} -- the maximum number of features to be used in the node;
\item[3.] \texttt{max\_depth} -- the maximum depth of the decision trees;
\item[4.] \texttt{min\_samples\_split} -- the minimum number of samples in the node of the decision tree required to make a split;
\item[5.] \texttt{min\_samples\_leaf} -- the minimum number of samples required to be in the leaf node (the end node in which the splitting finishes) of each tree;
\item[6.] \texttt{class\_weight} -- weights associated with each class (is required in the case of imbalance training sample)
\end{enumerate}

The number of estimators and the maximum depth (and/or minimum number of samples in the node) are mandatory hyperparameters. 
Moreover, a fine tuning of the parameters 3, 4 and 5 is crucial to avoid overfitting. 
For instance, setting the values of these parameters to their common values of $\{\infty, 2, 1\}$, respectively will most of the time lead to overfitting. 
In fact, if the depth of the decision tree is too high, and the minimum number of sample required to be in the leaf node is too small, then each single object in the training will have his own class characterized by its features. This will then make impossible to classify new, unknown objects, although the accuracy of classification of the training set will equals about 100\% \citep{Mansour1997}. 
Thus, to reduce overfitting, one has to limit the maximum depth of the decision trees (usually set to 3-20, depending on the amount and topology of the features), and/or the minimum number of samples in the nodes.

\subsection{Performance of the classification algorithms}
\label{sec:performance}
To directly compare the performance of the three algorithms that we tested, we use confusion matrices. These show the relative number of predicted objects in each of the three classes with respect to the number of true classes. 
The confusion matrices for RF, XGBoost and CatBoost are shown in Figure~\ref{fig:confmatr}. As we can see, RF provides the highest completeness for quasars ( $\approx 98.8\%$), but with a contamination of $\approx 0.8\%$ from stars and $\approx 0.1\%$ from galaxies; XGBoost shows the largest purity of quasar selection, but with a very low completeness (only $\approx 75.8\%$ quasars were classified as quasars). 

For the \texttt{GALAXY} class, CatBoost and RF provide very similar completeness (higher than XGBoost), with a very low contamination from star ($<0.1\%$) but CatBoost gives a much larger contamination from \texttt{QSO} ($\approx 0.25\%$).

As one can see, RF and CatBoost show very similar results. To better understand which was the best choice to make for our scientific purposes, we decided to compare the difference between the MCC value received for the training sample and the one received for the OOF sample, for each of the algorithms. Usually, a large difference between these two scores (training and validation) indicates an overfitting in the model, i.e. the classifier lose the generalization ability, and consequently is able to classify correctly only training data. 
For the RF, we received the following MCC values for training and OOF samples respectively: 0.9925 and 0.9892 (with a difference of $0.0033$). For CatBoost the MCC equals 0.9901 for training data and 0.9894 for the OOF sample, thus, in this case, the difference ($0.0007$) is almost 5 times smaller. 

Thus, the better ability to generalize good results on unseen dataset, combined with the fact that CatBoost keeps the purity and the completeness of the quasar selection at a very high level and it maximize the galaxies completeness, at the same time removing as much as possible stellar contaminants, convinced us that CatBoost is the optimal classifier for our purposes. 

\end{appendix}

\end{document}